\documentclass[preprint, superscriptaddress,amsmath, nofootinbib]{revtex4-1}
\usepackage{graphicx}
\usepackage{latexsym}
\usepackage{slashed}
\usepackage{bm}
\usepackage{color}
\usepackage{bbm}
\usepackage{amssymb}
\usepackage{amsmath}
\usepackage{mathrsfs}
\usepackage[colorlinks,citecolor=blue,urlcolor=blue,linkcolor=blue]{hyperref}
\usepackage{diagbox}

\usepackage{changes}

\def\keV{\,{\rm keV}}
\def\MeV{\,{\rm MeV}}
\def\GeV{\,{\rm GeV}}

\def\({\left(}
\def\){\right)}
\def\cm{{\,\rm cm}}

\def\beq{\begin{equation}}
\def\eeq{\end{equation}}

\begin{document}

\title{Spin-dependent sub-GeV Inelastic Dark Matter-electron scattering and Migdal effect: (I). Velocity Independent Operator}
	
\author{Jiwei Li}
\email{ljw@njnu.edu.cn}
\affiliation{Department of Physics and Institute of Theoretical Physics, Nanjing Normal University, Nanjing, 210023, China}

\author{Liangliang Su}
\email{liangliangsu@njnu.edu.cn}
\affiliation{Department of Physics and Institute of Theoretical Physics, Nanjing Normal University, Nanjing, 210023, China}

\author{Lei Wu}
\email{leiwu@njnu.edu.cn}
\affiliation{Department of Physics and Institute of Theoretical Physics, Nanjing Normal University, Nanjing, 210023, China}
	
\author{Bin Zhu}
\email{zhubin@mail.nankai.edu.cn}
\affiliation{Department of Physics, Yantai University, Yantai 264005, China}

\begin{abstract}

The ionization signal provide an important avenue of detecting light dark matter. In this work, we consider the sub-GeV inelastic dark matter and use the non-relativistic effective field theory (NR-EFT) to derive the constraints on the spin-dependent DM-electron scattering and DM-nucleus Migdal scattering. Since the recoil electron spectrum of sub-GeV DM is sensitive to tails of galactic DM velocity distributions, we also compare the bounds on corresponding scattering cross sections in Tsallis, Empirical and standard halo models. With the XENON1T data, we find that the exclusion limits of the DM-proton/neutron and DM-electron scattering cross sections for exothermic inelastic DM are much stronger that those for the endothermic inelastic DM. Each limits of the endothermic inelastic DM can differ by an order of magnitude at most in three considered DM velocity distributions.

\end{abstract}

\maketitle

\tableofcontents

\newpage
\section{Introduction}
Numerous astronomical and cosmological observations have provided evidence for the existence of dark matter (DM) in  the universe. However, besides its gravitational interaction, other physical properties of DM remain mystery. From the perspective of particle physics, dark matter may be made up of a hypothetical particle that is still undetected. Among the various conjectures, the weakly interacting massive particles (WIMPs) have been widely studied in the various experiments.

Direct detection that attempts to discern signals induced by DM at extremely low backgrounds has made great efforts in the past few  years~\cite{XENON:2018voc,XENON:2019gfn,COHERENT:2017ipa,XENON:2020rca,SuperCDMS:2018mne,Crisler:2018gci,PandaX-II:2018xpz,DarkSide:2018bpj,LUX:2018akb,SENSEI:2020dpa,EDELWEISS:2020fxc,EDELWEISS:2019vjv,LUX:2016ggv}. However, there is no any evidence of WIMP dark matter in the typical mass range. This strongly motivates the search for sub-GeV dark matter~\cite{PandaX:2022ood,SENSEI:2020dpa,EDELWEISS:2020fxc,EDELWEISS:2019vjv,LUX:2016ggv,Wang:2021nbf,Su:2020zny,Wang:2019jtk,DAMIC:2019dcn,Essig:2011nj,Dolan:2017xbu,LUX:2018akb,Vergados:2005dpd,GrillidiCortona:2020owp,Wang:2021oha,Bell:2021zkr,Bell:2019egg,Flambaum:2020xxo,Guo:2020oum,CDEX:2022kcd,An:2020tcg,Chao:2020yro,Liang:2018bdb,Ge:2022ius}. While the low momentum transfer of sub-GeV DM can not produce the observable nuclear recoil signal in the conventional detectors. With the improvements of direct detection experiments, we can access to the low mass DM by using the ionization events. Such signals can arise from the scattering of electrons with DM~\cite{DAMIC:2019dcn,PandaX:2022ood,SuperCDMS:2018mne,SENSEI:2020dpa,EDELWEISS:2020fxc,EDELWEISS:2019vjv,Essig:2011nj,LUX:2016ggv,Xia:2021vbz,An:2020tcg,Chao:2020yro,Liang:2018bdb,Guo:2020oum,Ge:2022ius,CDEX:2022kcd,He:2020wjs,Guo:2021vpb}, and the secondary effects in the DM-nuclear interactions, such as the Migdal scattering~\cite{Ibe:2017yqa,Dolan:2017xbu,LUX:2018akb,Vergados:2005dpd,GrillidiCortona:2020owp,Wang:2021oha,Bell:2021zkr,Bell:2019egg,Flambaum:2020xxo,Berghaus:2022pbu,Adams:2022zvg,Tomar:2022ofh,Blanco:2022pkt,He:2020sat}. There have been many studies on DM-electron scattering to date. For instance, in the context of elastic scattering, various operators for spin-dependent (SD) interactions are discussed in Ref.~\cite{Catena:2019gfa} in an effective field theory (EFT). The inelastic dark matter (iDM) model~\cite{Kopp:2016yji,Tucker-Smith:2001myb,Tucker-Smith:2004mxa,Gu:2022vgb,Duan:2018rls,Abdughani:2017dqs,Abdughani:2019wss,Finkbeiner:2007kk,Arina:2007tm,Chang:2008gd,Cui:2009xq,Fox:2010bu,Lin:2010sb,DeSimone:2010tf,Baryakhtar:2020rwy,Bramante:2020zos}, originally used to explain the DAMA anomaly, has also been used to study DM-electron scattering with spin-independent interactions to explain the XENON1T excess~\cite{Baryakhtar:2020rwy,Bramante:2020zos,He:2020wjs,Harigaya:2020ckz}. Dent et al.~\cite{Bell:2021zkr} showed some enlightening results on the Migdal effect of inelastic dark matter scattering with nuclei through the spin-independent (SI) interaction. However, there is still much scope for discussion of iDM-electron/Migdal scattering via SD interactions.

In this paper, we will study the ionization signals of sub-GeV inelastic dark matter (iDM), including Migdal effect and DM-electron scattering. Given the current strong constraints on the spin-independent (SI) cross section, we calculate the spin-dependent (SD) iDM-nucleus/electron scattering. We consider the Lagrangian density $\mathcal{L}_{\mathrm{int} }\supset  \bar{\chi}^{\,\prime}\gamma ^\mu  \gamma^5\chi\bar{\mathcal{N}}\gamma_\mu \gamma ^5\mathcal{N}$ for the axial-vector interaction of DM $\chi$ with the standard model particle $\mathcal{N}$ and derive the operator $\boldsymbol{O_4}=\vec{\mathbb{S}}_\chi\cdot \vec{\mathbb{S}}_\mathcal{N}$; this type of SD interaction is the only one in the leading order that is not suppressed by momentum transfer $\Vec{q}$. For some models, the SD interaction may still dominate, e.g. the scattering cross section for a Dirac DM particle interacting through its anomalous magnetic dipole moment, where the SD-like part (dipole-dipole) dominates in certain parameter space~\cite{Del_Nobile_2022}. Or when the DM is the Majorana fermion or a real vector boson, the SD interaction can naturally dominate (but is not always guaranteed)~\cite{Agrawal:2010fh}. In the future, if a signal associated with SD is observed, it will rule out the spinless DM particles by and large.

On the other hand, the velocity distribution function (VDF) of the local DM halo can have a non-negligible impact on the direct detection~\cite{Kuhlen:2009vh,McCabe:2010zh,Green:2017odb,Nunez-Castineyra:2019odi,Herrera:2021puj}. In particular, the electron recoil spectrum is sensitive to the high-velocity tail of the DM halo. As a benchmark distribution, the Standard Halo Model (SHM) is usually adopted~\cite{Drukier:1986tm}, however, it still can not accurately describe the distribution of DM in the Galaxy~\cite{Bozorgnia:2016ogo}. This motivates other alternative halo models for the VDF~\cite{Radick:2020qip}, such as Tsallis and Empirical models. We will also discuss their impacts on the exclusion limits of iDM-nucleus/electron scattering.

The paper is structured as follows. In Sec.~\ref{sec DMV}, we compare the velocity distribution functions for three models: the Standard Halo Model, the Tsallis model and the empirical model. In Sec.~\ref{idmn} and Sec.~\ref{idme}, we investigate the ionization rates of the spin-dependent scattering of the inelastic dark matter with the nucleus and the electron targets, respectively. With the available data, we obtain the exclusion limit for spin-dependent inelastic dark matter-nucleus Migdal/electron scattering in three velocity distribution models. Finally, we draw the conclusions in Sec.~\ref{sec6}.

\section{Dark Matter Velocity Distribution Function}
\label{sec DMV}

In the DM direct detections, the astrophysical properties of the local DM halo distribution, such as local DM density, mean DM velocity, etc., can significantly change the sensitivity. In particular, the electron spectrum is exceptionally sensitive to the high-velocity tail of the local velocity distribution of dark matter~\cite{Radick:2020qip,Herrera:2021puj,Maity:2020wic}. The most popular and widely used standard halo model (SHM) in DM direct detection experimental analysis, which assumes DM particles are in an isothermal sphere and obey the isotropic Maxwell-Boltzmann velocity distribution function (VDF). Although its simple analytical form is appealing~\cite{Bozorgnia:2016ogo}, this model cannot adequately explain the distribution of DM particles in the Galaxy. Consequently, it is important to investigate different velocity distribution models to substitute for the halo model. Based on the work in Ref.~\cite{Radick:2020qip}, this paper also introduces two additional velocity distribution models: Tsallis Model and an Empirical Model. We will discuss the effects on DM-Target scattering caused by different VDF models.

In the rest frame of the Galaxy, the SHM is given by
\begin{equation}
f_{\mathrm{SHM}}(\vec{v})\propto
\begin{cases}
\,e^{-|\vec{v}|^2/v^2_0}    &|\vec{v}|\leq v_{esc}\\
\,0&|\vec{v}|>v_{esc}.
\end{cases}
\end{equation}
The escape speed of the galaxy limits the speed of DM particles gravitationally bound to our galaxy, so a physical cut-off point is set at the local escape speed $v_{esc}$, with $v_0$ as the circular velocity at the Solar position~\cite{Kuhlen:2009vh}. The rotation curve in this model will be asymptotically flat at large $r$ (i.e. the distance from the centre of the Galaxy), and $v_0$ is usually regarded as the value of the curve at this point.
In the laboratory frame it has the following analytical forms
\begin{equation}
    f_{\mathrm{SHM}}(\vec v)=\frac{1}{K}e^{-|\vec v +\vec v_E|^2/v_0^2}\,\Theta(v_{\mathrm{esc}}-|\vec v +\vec v_E|),
\end{equation}
where $v_E$ is the Earth’s Galactic velocity. The velocity distribution of the SHM is truncated at the escape speed $v_{\mathrm{esc}}$ through the Heaviside function $\Theta$, with the normalization coefficient
\begin{equation}
    K=v_0^3 \left(\pi^{\frac{3}{2}}\,{\rm Erf}(\frac{v_{esc}}{v_0})-2\pi \frac{v_{esc}}{v_0}\mathrm{exp}\left(-\frac{v_{esc}^2}{v_0^2}\right)\right)
\end{equation}
that results from $\int f(\vec{v})\,\mathrm{d}^3 \vec{v}=1$.

The features of the local VDFs derived from DM cosmological simulations that include baryonic physics are largely consistent with the SHM; however, several studies ~\cite{Radick:2020qip,Kuhlen:2009vh,Vogelsberger:2008qb,Fairbairn:2008gz,March-Russell:2008rkh,Mao:2012hf} using data from DM-only simulations reveal a significant deviation from the overall trends manifested by the relevant local VDFs compared to the SHM. These simulations show that, especially in the high-velocity tail of the distribution, different features with the SHM will appear. One point worth making is that although adding baryons to the simulation makes the process more complex, it is nevertheless essential to restore the possible real universe.

Next, we discuss some alternative models in which the VDF of the Tsallis Model (Tsa)~\cite{Tsallis:1987eu} can be considered more compatible with the numerical results of $N-body$ simulations that include baryons~\cite{Hansen:2005yj,Ling:2009eh}. According to the statistical results of Tsallis, the definition of standard Boltzmann-Gibbs entropy is extended by introducing the entropy index $q_s$, as following
\begin{equation}
\begin{split}
    S_{q_s}&\equiv \frac{k}{q_s-1}\sum_{i=1}p_i\,(1-p^{q_s-1}_i)\\
    &=-k\sum_i\,p^{q_s}_i\ln_{q_s}p_i,
\end{split}
\label{entropy}
\end{equation}
where $p_i$ is the probability for a particle to be in state $i$, and $\ln_{q_s}p=(p^{1-q_s}-1)/(1-q_s)$. Note that $q_s$ is an arbitrary positive real number and that Eq.~\ref{entropy} recovers the standard Boltzmann-Gibbs entropy expression when the limit $q_s\to1$. Then, we can write down the velocity distribution function according to this Tsallis entropy
\begin{equation}
f_{\mathrm{Tsa}}(\vec{v})\propto
\begin{cases}
\,\left[1-(1-q_s)\dfrac{\vec v ^2}{v_0^2}\right]^{1/(1-q_s)}   & |\vec v|<v_{esc} \\
\,0         & |\vec v|\geq v_{esc}.
\end{cases}
\end{equation}
It is advantageous to use the Tsa model to elucidate the velocity distribution of the DM halo because the escape speed is already physically involved in the range $q_s<1$, determined by $v_{\mathrm{esc}}=v^2_0/(1-q_s)$, without the need for manual truncation, but the escape speed still needs to be set for $q_s>1$.
\begin{figure}[ht]
    \centering
    \includegraphics[height=8cm,width=8cm]{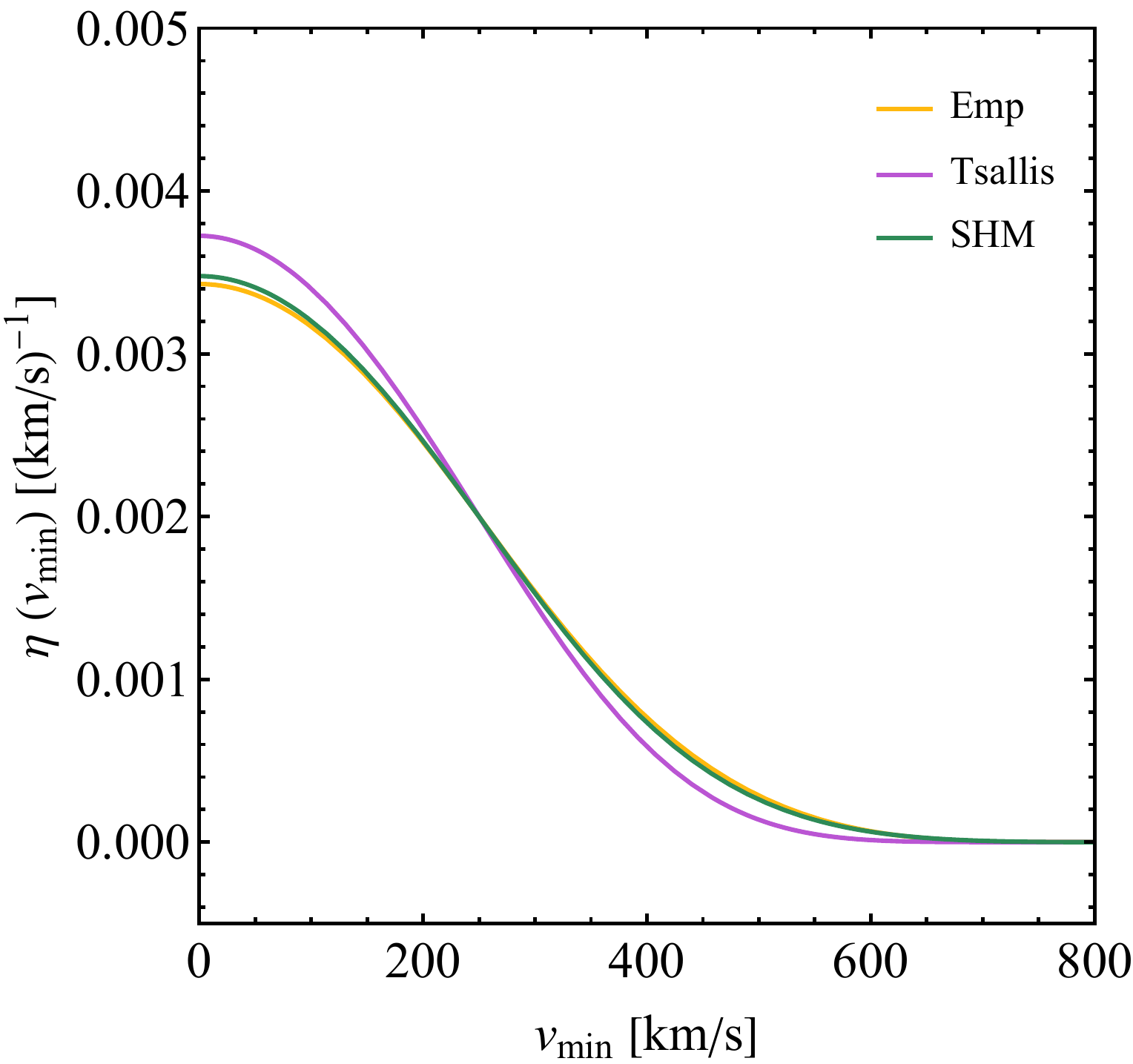}
    \caption{The $\eta(v_{min})$ is derived after integration over the velocity distribution, which varies with the parameters of the three models. For fixed the astrophysical parameters $v_0=238\,\rm{km/s}$, $v_E=250\,\rm{km/s}$ and $v_{esc}=544\,\rm{km/s}$, the orange line represents the Empirical model ($p=1.5$), the green represents the Standard Halo Model and the purple represents the Tsallis model ($q_s=0.809$).}
    \label{VDF}
\end{figure}
Finally, based on the work of Ref.~\cite{Mao:2012hf,Mao:2013nda,Radick:2020qip} another alternative model we introduce is an empirical model (Emp). It is derived from Hydrodynamical simulations with baryons on the data of DM-only cosmological simulation~\cite{Wu:2012wu,Klypin:2010qw}. In the Galactic rest frame, the empirical model described has a velocity distribution of the following form
\begin{equation}
    f_{\mathrm{Emp}}(\vec{v})\propto
\begin{cases}
\,\mathrm{exp}\left(-\dfrac{|\vec{v}|}{v_0}\right)\,(v_{esc}^2-|\vec v|^2)^p   & |\vec v|<v_{esc} \\
\,0         & |\vec v|\geq v_{esc}.
\end{cases}
\end{equation}
This empirical model is an exponential-based distribution, where $p$ is an adjustable parameter, and following the best-fit parameters for the Eris simulations~\cite{Radick:2020qip,Guedes:2011ux}, $p=1.5$ is set as our fiducial model. The shape of the VDF for this empirical model primarily relies on a proportional relationship, $r/r_s$, the ratio of the VDF's measured radius to the scaled radius of the halo density profile, and the uncertainty of the VDF is also derived from this quantity~\cite{Mao:2012hf}.

In Fig.~\ref{VDF}, we have depicted with solid lines of various colours the $\eta(v_{\rm{min}})$ resulting from the three velocity distribution models after integral $\eta(v_{\rm{min}})=\int \frac{\mathrm{d}^3 v}{v}\,f_\chi(v)\Theta(v-v_{\rm{min}})$. 
Here $v_{\rm{min}}$ is the minimum incoming DM velocity that causes nuclear recoil, and we will discuss it in the next section. In this paper, we adopted some astrophysical parameters suggested by recent work~\cite{Baxter:2021pqo},$v_0=238\,\rm{km/s}$~\cite{doi:10.1146/annurev-astro-081915-023441,abuter2021improved}, $v_{esc}=544\,\rm{km/s}$~\cite{Smith:2006ym} and $v_E=250\,\rm{km/s}$~\cite{Gelmini:2000dm} ( the Solar peculiar velocity from Ref.~\cite{10.1111/j.1365-2966.2010.16253.x} and average galactocentric Earth speed  from Ref.~\cite{McCabe:2013kea}.) corresponding to $q_s=0.809$ of the Tsallis model, and then compared the $\eta(v_{\rm{min}})$ values of different models with the same set of parameters. It can be observed that as the speed of DM shifts from low to high, the $\eta(v_{\rm{min}})$ transition in the Empirical model and the Standard Halo Model appears to be smoother, whereas the $\eta(v_{\rm{min}})$ of the Tsallis model is steeper than the other two models. In the low speed region, the $\eta(v_{\rm{min}})$ of the Emp and SHM diverge, although not significant (Tsa's diverges most from both). However as the DM speed increase, the Emp curve almost coincides with that of SHM.

In the following discussion, we turn our attention to inelastic dark matter-nucleus Migdal scattering and inelastic dark matter-electron scattering.
We will examine the impact of the velocity distribution model discussed above through the electron spectrum induced by these two processes.
\section{Inelastic Dark Matter-Nucleus Migdal Scattering}
\label{idmn}
We introduce a fermion dark matter $\chi$ with spin 1/2 coupling to a Standard Model (SM) particle $\mathcal{N}$~\cite{Dror:2019onn,Freytsis:2010ne,Del_Nobile_2022,Agrawal:2010fh,Dror:2019dib}. Assuming inelastic scattering between them, $\chi \mathcal{N}\to\chi^\prime \mathcal{N}$, mass splitting $\delta=m_{\chi^\prime}-m_{\chi}$ occurs between the incoming and outgoing dark matter (more details on kinematics are discussed below). If we consider that their interaction is via axial-vector-axial-vector couplings, the Lagrangian density $\mathcal{L}$ at low momentum transfer is
\begin{equation}
    \mathcal{L}_{\mathrm{int} }\supset  \bar{\chi}^{\,\prime}\gamma ^\mu  \gamma^5\chi\bar{\mathcal{N}}\gamma_\mu \gamma ^5\mathcal{N}. 
\end{equation} 
This is called the standard spin-dependent interaction in non-relativistic effective field theory (NR-EFT) and is usually reduced to the type of the two spin operators, $-4\vec{\mathbb{S}}_\chi\cdot \vec{\mathbb{S}}_\mathcal{N}=-4\boldsymbol{O}_4$. Such a spin-dependent interaction is the only one in the leading order not suppressed by momentum transfer $\Vec{q}$. This may allow us to place stronger constraints on the DM-nucleus scattering of SD interactions.

\subsection{Calculations}
We begin with the perspective of inelastic dark matter-nucleon scattering kinematics.
In general, there are two different ways to reveal inelasticity~\cite{Del_Nobile_2022}, DM particle of mass $m_{\chi}$ undergo mass splitting after scattering with nucleus become to $m_{\chi^\prime}-m_{\chi}=\delta$, or there is the possibility of the nucleon transitioning from a low-energy state to an excited state. The latter case has been studied in many literatures~\cite{Ellis:1988nb,Vergados:2003st,Engel:1999kv,Vergados:2013raa,Baudis:2013bba,McCabe:2015eia,Vietze:2014vsa}, and for the sake of simplicity we do not consider this possibility in this work.

We focus on the process $\chi(\vec{p})+N(\vec{k}) \to \chi'(\vec{p}^{\,'})+N(\vec{k}^{\,'})$, where $\chi$ and $\chi'$ are dark matter particles in the initial and final states, respectively, and $N$ is a nucleon. For non-relativistic limit, inspired by the conservation of energy in the center-of-mass (CM) framework we have
\begin{equation}
\begin{split}
    \frac{1}{2}\mu_N v^2&=\frac{p^{\prime\,2}}{2m_{\chi'}}+\frac{k^{\prime\,2}}{2m_{N}}+\Delta\\
    &=\frac{(\vec{p}+\vec{q})^2}{2m_{\chi'}}+\frac{(\vec{k}-\vec{q})^2}{2m_{N}}+\Delta,
\end{split}
\label{energy}
\end{equation}
where $\mu_N=\frac{m_\chi m_N}{(m_\chi+m_N)}$ is the reduced mass of the initial $\chi-N$ system, $\vec{v}\equiv\frac{\vec{p}}{m_{\chi}}-\frac{\vec{k}}{m_N}$ is the relative velocity between the DM particle and the nucleon.
The momentum transfer $\vec{q} \equiv \vec{p}-\vec{p'}=\vec{k'}-\vec{k}$. It should be noted that the momentum transfer $\vec{q}$ is approximately Galilean invariant in the inelastic scattering under NR boost when the mass splitting $\left| \delta \right|\ll m_\chi$.
$\Delta$ shows the initial kinetic energy lost due to inelastic effects.
To better discuss the Migdal effect and electron scattering that follow, we write $\Delta$ here as $\Delta=E_{\mathrm{em}}+\delta$ (for nucleus
scattering, $\Delta=\delta$) , and $E_{\mathrm{em}}$ is the electromagnetic energy available to excite the electron.
In this paper we have conventionally defined that $\delta>0$ corresponds to the endothermic scattering, while $\delta<0$ is the exothermic scattering. Comparing the $\Delta$ with initial kinetic energy $E_{\mathrm{kin}}$ of the system, $\Delta=0$ corresponds to the usual elastic scattering. Apparently, we can know from Eq.~\ref{energy} that the maximum possible value of $\Delta$ for the scattering should be equal to the initial available kinetic energy
\begin{equation}
    E_{\mathrm{kin}}\equiv\frac{1}{2}\mu_N v^2=\Delta_{max}.
\label{maxdelta}
\end{equation}
Significantly, the masses of DM and nucleons are so large compared to kinetic energy that scattering is only kinematically allowed in the $|\Delta| \ll m_\chi$ scenario.

To facilitate our calculations, we set $\vec{p}_i\equiv\mu_N\vec{v}$ for initial system momentum, and in the final system $\chi'-N$, we can write $E_{f}=\frac{1}{2}\mu'v'^2+\Delta$.
In the NR limit approximation, we will take $\Delta/\mu^{(\prime)}_N$ as a parameter of order $\mathcal{O}(v^2)$, thus we have
\begin{equation}
    v^{\prime\,2}=v^2-2\frac{\Delta}{\mu_N}
\end{equation}
and the square of the momentum of the final system $p^{\prime\,2}_f$
\begin{equation}
    p^{\prime\,2}_f=\mu^{\prime\,2}_N v^{\prime\,2}\simeq\mu^2_N v^2-2\mu_N\Delta.
\label{pf}
\end{equation}
The transfer momentum $\vec{q}=\left(\vec{p}^{\,\prime}_f-\vec{p}_i\right)$ is the same in both frames, so that we then can express the atomic recoil energy in the frame of the detector as
\begin{equation}
    E_R=\frac{\mu^2_Nv^2}{m_N}\left(1-\cos{\theta}\sqrt{1-\frac{2\Delta}{\mu_Nv^2}}\right)-\frac{\mu_N\Delta}{m_N},
\label{ER}
\end{equation}
where $\theta$ is the DM-nucleon scattering angle in the CM frame. 
It is worth mentioning that the derivation above for $\mu_N\sim\mathcal{O}(\GeV)$ , if $\Delta\sim\mathcal{O}(\keV)$, there is $\mu_N\simeq\mu^\prime_N=m^\prime_\chi m_N/(m^\prime_\chi+m_N)$. But in numerical calculation, we still maintain the complete expansion $(\mu_N\neq\mu^\prime_N)$. 

We can also see from the Eq.~\ref{ER} that if the incoming DM has a fixed speed, there will be a maximum $E_{R}^{\mathrm{max}}$ and a minimum $E_{R}^{\mathrm{min}}$ of the recoil energy, corresponding to $\theta=\pi$ and $0$, respectively. Likewise, when the DM particle imparts a given recoil energy to the target nucleus, the incident speed of DM is kinematically limited.
If we express the momentum transfer $q=\sqrt{2m_NE_R}$ in terms of energy recoil, then we get the minimum DM velocity that can cause nuclear recoil,
\begin{equation}
    v_{\mathrm{min}}(E_R)=\left\vert\frac{q}{2\mu_N}+\frac{\Delta}{q}\right\vert=\frac{1}{\mu_N\sqrt{2m_NE_R}}\left\vert m_NE_R+\mu_N\Delta \right\vert.
\label{Vmin}
\end{equation}

Next, we will introduce a non-relativistic effective field theory to help us discuss inelastic dark matter-nucleus scattering. Given the average velocity of DM in the galactic halo is $v\sim\mathcal{O}(10^{-3})$, the non-relativistic effective field theory provides a bottom-up framework to study the DM direction detection~\cite{Fan:2010gt,Fitzpatrick:2012ix,Fitzpatrick:2012ib,Anand:2013yka,Bell:2019egg,Liu:2017kmx,Gondolo:2020wge,Gondolo:2021fqo}. This formalism enables the decomposition of the interaction of dark matter with the nucleus into two classes of response functions. And it allows us to use pre-calculated nuclear form factors for the relevant interaction operators.

According to the work of Haxton et al. \cite{Fitzpatrick:2012ix,Fitzpatrick:2012ib,Anand:2013yka}, they established an EFT based on elastic DM-nucleus scattering. This approach allows to construct a series of effective operators from four Galilean invariants: the DM particle spin $\vec{\mathbb{S}}_\chi$, nucleon spin $\vec{\mathbb{S}}_N$, the momentum transfer $i\vec{q}$ and the transverse velocity $\vec{v}^{\perp}_{el}\equiv\vec{v}+\frac{\vec{q}}{2\mu_N}$. However, in the case of inelastic scattering, it is necessary to modify the quantity due to mass splitting.
As indicated by the formalism in the Ref.~\cite{Barello:2014uda}, the NR-EFT for inelastic scattering of dark matter is a direct extension of elastic scattering. 
It pointed out in the context that at the leading order of the $v$ expansion, the only modification made is that $\vec{v}^\perp_{el}$ changed from elastic scattering. According to the kinematics of inelastic scattering, the mass splitting $\delta$ allows for a contribution to the incident velocity component perpendicular to the momentum transfer $\vec{q}$, so that a new Galilean invariant on inelastic scattering can be obtained by adding a new component for modification
\begin{equation}
    \vec{v}^\perp_{el}\to\vec{v}^\perp_{inel}\equiv\vec{v}+\frac{\vec{q}}{2\mu_N}+\frac{\Delta}{\left\vert\,\vec{q}\,\right\vert^2}\,\vec{q}.
\label{Ninel}
\end{equation}
The above equation satisfies $\vec{q}\cdot\vec{v}^\perp_{inel}=0$ due to the conservation of energy. The effect of this inelasticity will be directly reflected in the DM particle response function $\mathcal{R}^{\tau\tau^\prime}_\mathrm{X}$ rather than the nucleon response function $\mathcal{W}^{\tau\tau^\prime}_\mathrm{X}$. However, we are concerned with the effective spin-dependent operator $\boldsymbol{O}_4=\vec{\mathbb{S}}_\chi\cdot \vec{\mathbb{S}}_\mathcal{N}$, which does not depend on $\vec{v}^\perp_{inel}$. For our calculations, we can still use the nucleon matrix elements from Ref.~\cite{Anand:2013yka}.

For a given Lagrangian, the invariant amplitude of the DM-nucleon can be obtained using spherical harmonics and multipole expansions,
\begin{equation}
\begin{split}
     \mathcal{M}&=\sum_{\tau=0,1}\langle j_\chi^\prime,M_\chi^\prime; j_N^\prime,M_N^\prime|\,\boldsymbol{O}_{JM;\tau}\left(q\right)\,|j_\chi,M_\chi;j_N,M_N\rangle\\
     &\equiv\sum_{\tau=0,1}\langle j_\chi^\prime,M_\chi^\prime; j_N^\prime,M_N^\prime|\,\sum^A_{i=1}\boldsymbol{O}_{JM}\left(q\vec{x}_i\right)t^\tau\left(i\right)\,|j_\chi,M_\chi;j_N,M_N\rangle.
\end{split}
\end{equation}
Here $\boldsymbol{O}_{JM;\tau}\left(q\right)$ contains six operators familiar to the standard model electroweak interaction theory: $\boldsymbol{M}$, $\boldsymbol{\Sigma}^\prime$, $\boldsymbol{\Delta}$, $\boldsymbol{\Sigma}^{\prime\prime}$, $\boldsymbol{\Phi}^{\prime\prime}$, $\Tilde{\Phi}^\prime$. This is the result obtained by considering only elastic transitions and assuming that the nuclear ground state obeys CP and parity conservation. According to semi-leptonic electroweak theory~\cite{DONNELLY19791,DONNELLY1979103,walecka2004theoretical}, the only spin-dependent interactions of interest to us are only two related single particle operators, $\boldsymbol{\Sigma}^\prime$ and $\boldsymbol{\Sigma}^{\prime\prime}$, corresponding to axial transverse and axial longitudinal operators, respectively,
\begin{equation}
\begin{split}
    \boldsymbol{\Sigma}^\prime_{JM;\tau}\left(q^2\right)&\equiv-i\sum_{i=1}^A \left\{ \frac{1}{q}\vec{\nabla}_i\times\vec{\boldsymbol{M}}^M_{JJ}\left(q\vec{x}_i\right)\right\}\cdot\vec{\sigma}\left(i\right)\tau_3\left(i\right)\\
    \boldsymbol{\Sigma}^{\prime\prime}_{JM;\tau}\left(q^2\right)&\equiv\sum^A_{i=1}\left\{\frac{1}{q}\vec{\nabla}_i\boldsymbol{M}_{JM}\left(q\vec{x}_i\right)\right\}\cdot\vec{\sigma}\left(i\right)\tau_3\left(i\right).
\end{split}
\end{equation}
By averaging over initial spins and summing over outgoing spins, we then write down the DM-nucleus scattering transition probability,
\begin{equation}
\begin{split}
    P_{tot}&=\frac{1}{2j_\chi+1}\frac{1}{2j_N+1}\sum_{\mathrm{spins}}\left\vert  \mathcal{M} \right\vert^2\\
    &=\frac{4\pi}{2j_N+1}\sum_{\tau=0,1}\sum_{\tau^\prime=0,1}
    \left[\mathcal{R}^{\tau\tau^\prime}_{\boldsymbol{\Sigma}^{\prime\prime}}\left(\vec{v}_{T_{inel}}^{\perp\,2},\frac{\vec{q}^{\,2}}{m_N^2}\right)\mathcal{W}^{\tau\tau^\prime}_{\boldsymbol{\Sigma}^{\prime\prime}}\left(q^2\right)+\mathcal{R}^{\tau\tau^\prime}_{\boldsymbol{\Sigma}^{\prime}}\left(\vec{v}_{T_{inel}}^{\perp\,2},\frac{\vec{q}^{\,2}}{m_N^2}\right)\mathcal{W}^{\tau\tau^\prime}_{\boldsymbol{\Sigma}^{\prime}}\left(q^2\right)\right],
\end{split}
\label{Respon}
\end{equation}
where $j_\chi$ and $j_N$ label the dark matter and nuclear spin, respectively.
The Eq.~\ref{Respon} expresses the transition probability as the product of the DM particle response functions $\mathcal{R}^{\tau\tau^\prime}_\mathrm{X}$ and nuclear response functions $\mathcal{W}^{\tau\tau^\prime}_\mathrm{X}$. The former is determined by the bilinear functions $c^\tau_i$'s in the EFT coefficients, which distinguishes particle physics well from nuclear physics. In the isospin basis $c^{\tau(\prime)}_i$, here we list the DM particle response functions considered,
\begin{equation}
\begin{split}
  \mathcal{R}^{\tau\tau^\prime}_{\boldsymbol{\Sigma}^\prime}\left(\vec{v}_{T_{inel}}^{\perp\,2},\frac{\vec{q}^{\,2}}{m_N^2}\right)
  &=\frac{1}{8}\left[\frac{\vec{q}^{\,2}}{m_N^2}\vec{v}^{\perp\,2}_{T_{inel}}c^\tau_3 c^{\tau^\prime}_3+\vec{v}^{\perp\,2}_{T_{inel}}c^\tau_7 c^{\tau^\prime}_7\right]+\frac{j_\chi(j_\chi+1)}{12}\left[c^\tau_4 c^{\tau^\prime}_4\right.\\
  &+\left.\frac{\vec{q}^{\,2}}{m_N^2}c^\tau_{9} c^{\tau^\prime}_{9}+\frac{\vec{v}_{T_{inel}}^{\perp\,2}}{2}\left(c^\tau_{12}-\frac{\vec{q}^{\,2}}{m_N^2}c^\tau_{15}\right)\left(c^{\tau\prime}_{12}-\frac{\vec{q}^{\,2}}{m_N^2}c^{\tau^\prime}_{15}\right)+\frac{\vec{q}^{\,2}}{2m_N^2}\vec{v}^{\perp\,2}_{T_{inel}}c^\tau_{14} c^{\tau^\prime}_{14}\right]\\\\
  \mathcal{R}^{\tau\tau^\prime}_{\boldsymbol{\Sigma}^{\prime\prime}}\left(\vec{v}_{T_{inel}}^{\perp\,2},\frac{\vec{q}^{\,2}}{m_N^2}\right)
  &=\frac{\vec{q}^{\,2}}{4m_N^2}c^\tau_{10} c^{\tau^\prime}_{10}+\frac{j_\chi(j_\chi+1)}{12}\left[c^\tau_4 c^{\tau^\prime}_4\right.\\
  &+\left.\frac{\vec{q}^{\,2}}{m_N^2}\left(c^\tau_{4} c^{\tau^\prime}_{6}+c^\tau_{6} c^{\tau^\prime}_{4}\right)+\frac{\vec{q}^{\,4}}{m_N^4}c^\tau_{6} c^{\tau^\prime}_{6}+\vec{v}^{\perp\,2}_{T_{inel}}c^\tau_{12} c^{\tau^\prime}_{12}+\frac{\vec{q}^{\,2}}{m_N^2}\vec{v}^{\perp\,2}_{T_{inel}}c^\tau_{13} c^{\tau^\prime}_{13}\right].
\end{split}
\label{DM Rs}
\end{equation}
The nuclear response functions $\mathcal{W}^{\tau\tau^\prime}_\mathrm{X}$, obtained by multipoles expansion and summing over the nuclear states,
\begin{equation}
\begin{split}
    \mathcal{W}_{\boldsymbol{\Sigma}^{\prime\prime}}^{\tau\tau^\prime}(q^2)&=
    \sum^{\infty}_{J=1,3,\ldots}\langle j_N \left \|\,\boldsymbol{\Sigma}^{\prime\prime}_{J;\tau}(q)\,\right\|j_N\rangle\langle j_N \left \|\,\boldsymbol{\Sigma}^{\prime\prime}_{J;\tau^\prime}(q)\,\right\|j_N\rangle\\
    \mathcal{W}_{\boldsymbol{\Sigma}^{\prime}}^{\tau\tau^\prime}(q^2)&=
    \sum^{\infty}_{J=1,3,\ldots}\langle j_N \left \|\,\boldsymbol{\Sigma}^{\prime}_{J;\tau}(q)\,\right\|j_N\rangle\langle j_N \left \|\,\boldsymbol{\Sigma}^{\prime}_{J;\tau^\prime}(q)\,\right\|j_N\rangle
\end{split}
\label{nuclear Rs}
\end{equation}
where $\mathcal{W}_{\boldsymbol{\Sigma}^{\prime}}^{\tau\tau^\prime}$ and $\mathcal{W}_{\boldsymbol{\Sigma}^{\prime\prime}}^{\tau\tau^\prime}$ only receive contributions from the odd multipoles. A more complete formulation of Eq.~\ref{Respon}, Eq.~\ref{DM Rs} and Eq.~\ref{nuclear Rs} is shown in the Ref.~\cite{Anand:2013yka}. The full amplitude 
or the nuclear responses can be calculated using the package {\it DMFormFactor}. Notice that, relativistic normalisation is used in Eq.~\ref{Respon} to produce a dimensionless $\left\vert  \mathcal{M} \right\vert^2$, which is achieved by multiplying by a factor of $\left(4m_\chi m_T\right)^2$. From the transition probability $P_{tot}$, one can immediately obtain the differential cross section
\begin{equation}
    \frac{\rm{d}\sigma}{\mathrm{d}E_R}=\frac{2m_T}{4\pi v^2}\left[\frac{1}{2j_\chi+1}\frac{1}{2j_N+1}\sum_{\mathrm{spins}}\left\vert  \mathcal{M} \right\vert^2\right].
\end{equation}

Next, we turn our attention to the Migdal effect of iDM-nucleus scattering. Based on the work in Ref.~\cite{Ibe:2017yqa,Essig:2019xkx,Baxter:2019pnz,Berghaus:2022pbu,Adams:2022zvg,Tomar:2022ofh,Blanco:2022pkt,Liang:2018bdb,Cox:2022ekg}, we will briefly review the Migdal effect and present the main formulas to facilitate our calculation of the scattering cross section. 
The Migdal effect is the process of atomic ionization or excitation. In the scattering of DM particles and nuclear, the nucleus suddenly receives a transfer momentum $\vec{q}$, and the electron cloud cannot `catch up' instantaneously, which makes it possible to detect the subsequent electromagnetic signatures. 
Thus, the theoretical calculation of the Migdal effect and the DM-electron scattering rate is closely related.

In kinematics, we can obtain the formulae of the Migdal scattering by replacing the electron mass $m_e$ with the nucleus mass $m_N$ in those of the DM-electron scattering.
To demonstrate the physical process of Migdal, following Ref.~\cite{Ibe:2017yqa}, it is assumed that both the incoming and outgoing DM are plane waves. However, the outgoing atom is regarded as an atom in an excited state, where the ionized electrons belong to the continuum of the atomic Hamiltonian~\cite{Baxter:2019pnz}. In this formalism, treating the nucleus and electron as a single many-particle system would allow us to treat the transfer momentum $\vec{q}$ as originating from the DM rather than other specific components.
According to the conservation of energy in Eq.~\ref{energy}, we have
\begin{equation}
     E_{\mathrm{em}}=\vec{q}\cdot\vec{v}+\frac{q^2}{2\mu_N}-\delta,
\label{N}
\end{equation}
where $E_{\mathrm{em}}=E_{e,f}-E_{e,i}$ is the transfer energy available for scattered electrons. There exists a maximum value of $E_{\mathrm{em}}$, which can be derived from Eq.~\ref{maxdelta},
\begin{equation}
    E^{max}_{\mathrm{em}}\simeq\frac{1}{2}m_\chi v^2_{max}-\delta,
\label{maxEem}
\end{equation}
where assumed $m_\chi\ll m_N$, $v_{max}$ is the maximum DM incoming velocity (in laboratory frame). This indicates that the maximum value $E^{max}_{\mathrm{em}}$ is not related to the initial occupied energy level of Migdal electrons and target nucleus. 
We would like to point out that these inelastic effects mainly affect in kinematics.

Nevertheless, the dynamics of Migdal and electron scattering differ significantly depending on whether the DM interacts directly with electrons or the nucleus. To clarify their connection, we briefly review the process from isolated atom reduction to nuclear recoil and projection onto the electron cloud~\cite{Ibe:2017yqa}. In the relativistic limit, we convert the dark matter-nucleus interaction into an interaction potential $V_{\mathrm{int}}$, then the total Hamiltonian for the atom can be written as
\begin{equation}
    H_{\mathrm{tot}}=H_\mathrm{A}+\frac{\hat{p}^2_\chi}{2m_\chi}+V_{\mathrm{int}}(\vec{x}_N-\vec{x}_\chi),
\end{equation}
where $\vec{x}_N$ and $\vec{x}_\chi$ represent the positional operators of nucleus and DM, and $H_\mathrm{A}$ is the approximate Hamiltonian of the atomic system. Therefore, the elements of the transition matrix are derived by using reduced atomic eigenstates of $H_\mathrm{tot}$
\begin{equation}
    iT_{FI}\thicksim F(q^2_N)\times \mathcal{M}(q^2_N) \times Z_{FI}(q_e)\times i(2\pi)^4\delta^4(p_F-p_I).
\label{T}
\end{equation}
Here $T_{FI}$ is decomposed into the nuclear form factor $F(q^2_N)$ and the DM-nucleon scattering invariant matrix element $\mathcal{M}(q^2_N)$, both evaluate the interaction of nucleons. And the factor $Z_{FI}$ associated with the electron cloud transition. This treatment makes explicit the conservation of momentum-energy at invariant amplitude. 

Notice that Eq.~\ref{T} assumes that the initial state of atoms in the laboratory frame are stationary, i.e. $\vec{v}_I=0$. Moreover, $V_{\mathrm{int}}$ is the interaction potential between the nucleus and the DM, it does not contain the position operator $\vec{x}$ of the electron, so theoretically, the electron cannot be induced to transition. 
Assume that momentum $\vec{q}$ is transferred instantaneously to the nucleus, in which case the entire atom suddenly obtains velocity $\vec{v}_A\equiv\frac{\vec{q}_e}{m_e}$ and leaves its stationary electrostatic potential.
At this moment the wave function of the electron of the moving atom will change, considering $\vec{q}_e\equiv\frac{m_e}{m_N}\vec{q}$ as the effective momentum of the electron. Following the method of Ref.~\cite{Ibe:2017yqa}, where electron transitions and nucleon scattering are linked to construct the approximate energy eigenstates of the moving atoms by applying the Galilean transformation with the velocity parameter $\vec{v}_A$.

Now we will assess the factor $Z_{FI}$,  which is the sum of three probabilities as given in the Ref.~\cite{Ibe:2017yqa},
\begin{equation}
    \sum_F |Z_{FI}|^2=|Z_{II}|^2+|Z_{\mathrm{exc}}|^2+|Z_{\mathrm{ion}}|^2.
\end{equation}
Here $|Z_{II}|^2$ represents the probability that the electron is unaffected by the nuclear recoil (mention that this is the result in $\mathcal{O}(\frac{q^2_e}{\langle r\rangle^2})$.), whereas $|Z_{\mathrm{exc}}|^2$ and $|Z_{\mathrm{ion}}|^2$ denote the probabilities of electron excitation and ionization, respectively. 
The ionization factor $Z_{\mathrm{ion}}(q_e)$ involve
\begin{equation}
    Z_{\mathrm{ion}}(q_e)=\langle F|e^{i\frac{m_e}{m_N}\vec{q}\cdot\sum_{\zeta}\vec{x}^{(\zeta)}}|I\rangle\thicksim\sum^{\mathrm{\zeta}}\langle f|\,i\vec{q}_e\cdot\vec{x}^{\,(\zeta)}|i\rangle,
\label{ion}
\end{equation}
where $m_e$ and $m_N$ are the mass of the electron and nucleus, respectively. The last term of the above equation considers the leading order of the Taylor expansion of $\vec{q}_e$.
We have made approximations by factoring the wave functions of the initial and final electron clouds, $|I\rangle$ and $|F\rangle$, so that only a single electron (with the position operator $\vec{x}^{\,(\zeta)}$, i.e. $\vec{x}^{\,(\zeta)}$ denotes the position of $\zeta^{th}$ electron in electron cloud.) involved in the transition between the single-electron states $|i\rangle$ and $|f\rangle$. 

Furthermore, we will quickly write down the single-electron transition amplitude for the direct interaction of DM with the electron at coordinate $\vec{x}^{\,(\eta)}$,
\begin{equation}
    \langle F|e^{i\vec{q}\cdot\sum_{\eta}\vec{x}^{\,(\eta)}}|I\rangle\thicksim\sum^{\eta}\langle f|\,i\vec{q}\cdot\vec{x}^{\,(\eta)}|i\rangle.
\label{electronTrans}
\end{equation}
Comparing Eq.~\ref{ion} and Eq.~\ref{electronTrans}, we can see that the Migdal effect and DM-electron scattering are very similar in form, while the critical difference between them is the transfer momentum $\vec{q}_e$ and $\vec{q}$.
For the latter, the electrons directly obtain the momentum lost by the DM. The transfer momentum $\vec{q}_e$ received by the electrons in the Migdal process is suppressed by a factor of $10^{-3}/\mathrm{A}$ (here $\mathrm{A}$ is atomic mass number).

To calculate the electron ionization probability of the Migdal effect and electron scattering in isolated atoms, we rely on the work in the Ref.~\cite{Essig:2011nj,Essig:2012yx,Essig:2019xkx} to establish their precise relationship. From the dimensionless ionization form factor $\left\vert f^{nl}_{ion}(k_e,q)\right\vert^2$ defined in the Ref.~\cite{Essig:2012yx,Essig:2011nj}, we can rewrite the Eq.~\ref{electronTrans} as
\begin{equation}
    |Z_{\mathrm{ion}}|^2\equiv\left\vert f^{nl}_{ion}(k_e,q)\right\vert^2=\frac{2k_e}{8\pi^3}\times\sum_{\substack{\mathrm{occupied}\\\mathrm{states}}}\sum_{l^\prime\,m^\prime}\left\vert\langle k_e,l^\prime,m^\prime|\,e^{i\vec{q}\cdot\vec{x}}\,|n,l\rangle\right\vert^2.
\end{equation}
This represents the sum over final state angular variables $l^\prime$, $m^\prime$  and degenerate, occupied initial states. The initial state wave function of bound electrons in isolated atoms is characterized by the principal quantum number $n$ and the angular momentum quantum number $l$, and the final state is a continuous unbound electron state with momentum $k_e=\sqrt{2m_eE_e}$, which represents the quantum numbers are $l^\prime$ and $m^\prime$. 
We adopt the ionization form factor $\left\vert f^{nl}_{ion}(k_e,q)\right\vert^2$ given in the work~\cite{Hamaide:2021hlp} to derive our results. However, the ionization form factor provided in their research does not adequately describe the ionization behavior of transfer momentum $\vec{q}$ below $1\keV$, hence we employ dipole approximation to extend it,
\begin{equation}
    \left\vert f^{nl}_{ion}(k_e,q)\right\vert^2=\left(\frac{q}{q_0}\right)^2\times\left\vert f^{nl}_{ion}(k_e,q_0)\right\vert^2.
\end{equation}
For the xenon atom, $q_0\lesssim1\keV$ is usually chosen so that the above approximation holds.

According to the previous description and the Ref.~\cite{Essig:2019xkx}, we utilize the parameter $q_e$ to characterize the ionization probability of Midgal, it can be expressed as
\begin{equation}
    \sum_{n,l}\frac{\mathrm{d}}{\mathrm{d}\ln E_{e}}p^c_{q_e}(n,l\to E_e)=\frac{\pi}{2}\left\vert f^{nl}_{ion}(k_e,q_e)\right\vert^2.
\end{equation} 
Thus we can derive the ionization differential event rates induced by the Migdal effect in iDM-nucleus scattering,
\begin{equation}
    \frac{\mathrm{d}R}{\mathrm{d}E_R\mathrm{d}E_{\mathrm{em}}\mathrm{d}v}\simeq\frac{\mathrm{d}R_0}{\mathrm{d}E_R\mathrm{d}v}\times\frac{1}{2\pi}\sum_{n,l}\frac{\mathrm{d}}{\mathrm{d}E_{e}}p^c_{q_e}(n,l\to E_e),
\label{totrate}
\end{equation}
where
\begin{equation}
    \frac{\mathrm{d}R_0}{\mathrm{d}E_R}=N_T\frac{\rho_\chi}{m_\chi}\int_{v>v_{\mathrm{min}}}\frac{\mathrm{d}\sigma}{\mathrm{d}E_R}vf(v)\mathrm{d}^3v.
\label{rate}
\end{equation}
This is decomposed into the standard elastic DM-nucleus scattering differential rate $\frac{dR_0}{dE_R}$ multiplied by the electron ionization probability, where $f(v)$ is the local velocity distribution function of the DM, $\rho_\chi$ is the local DM density (take $\rho_\chi\simeq0.3\GeV/\mathrm{cm}^3$ in our calculations ), $N_T$ is the number density of target nuclei in the detector. The total electromagnetic energy $E_{\rm{em}}$ is defined as the sum of the outgoing unbound electron energy $E_e$ and the binding energy $E_{nl}$ between the corresponding levels: $E_{\rm{em}}=E_e+E_{nl}$.
For Migdal scattering, the electron equivalent energy detected by the detector is given by
\begin{equation}
    E_{\mathrm{det}}=\mathcal{Q}\,E_R+E_{\mathrm{em}}=\mathcal{Q}\,E_R+E_e+E_{nl},
\end{equation}
where $\mathcal{Q}$ is the introduced quenching factor that depends on the nuclear recoil energy. The quenching factor of different target nucleus will also be different, and there have been a series of measurement results for xenon~\cite{Dahl:2009nta,XENON100:2011uwh}; here, we take a fixed $\mathcal{Q}=0.15$~\cite{Ibe:2017yqa}. Finally we can obtain the detection energy spectrum,
\begin{equation}
    \frac{\mathrm{d}R}{\mathrm{d}E_{\mathrm{det}}\,\mathrm{d}v}\simeq\int \mathrm{d}E_R\,\mathrm{d}E_{\mathrm{em}}\frac{\mathrm{d}R}{\mathrm{d}E_R\,\mathrm{d}E_{\mathrm{em}}\,\mathrm{d}v}\times\delta(E_{\mathrm{det}}-\mathcal{Q}E_R-E_{\mathrm{em}}).
\end{equation}

\subsection{Numerical Results and Discussions}

To facilitate comparison with other results, we set the cross section of the DM-nucleon at transfer momentum $q=0$ as
\begin{equation}
    \Bar{\sigma}_{\chi n}(q=0)\equiv\frac{c_i^{n\,2}\mu^2_{\chi n}}{\pi}.
\end{equation}
To compensate for the dimension of the coefficient $c_i^n\,(\mathrm{Energy})^{-2}$, we have maintained the convention of multiplying by the square of electroweak interaction strength $m^{-2}_v  \equiv (246.2\GeV)^{-2} $ in our calculations.

\begin{figure}[ht]
\centering
\includegraphics[height=8cm,width=7.5cm]{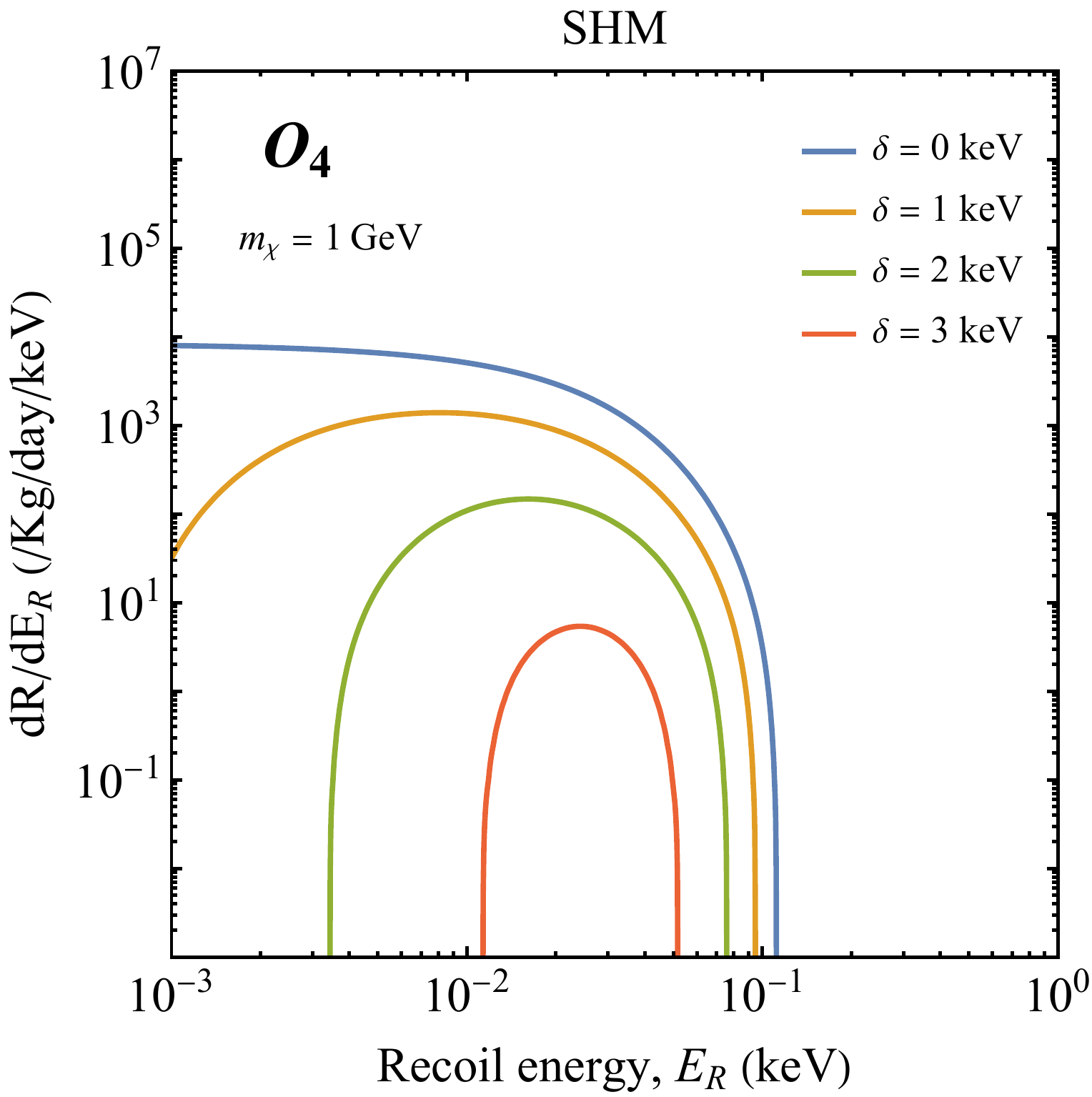}
\hspace{1cm}
\includegraphics[height=8cm,width=7.5cm]{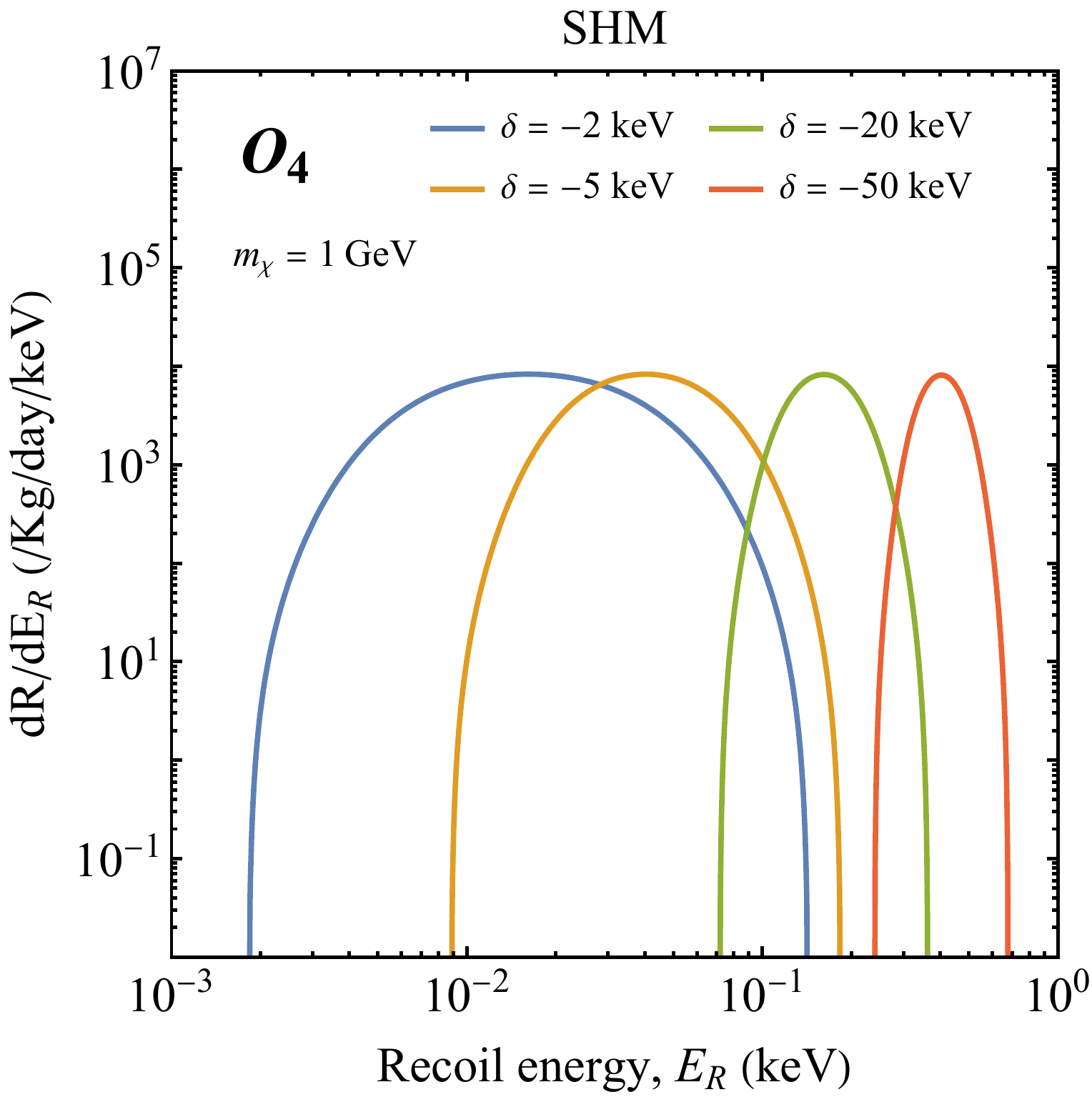}
\caption{The nuclear recoil spectrum is derived from the inelastic dark matter scattered with xenon atoms in the Standard Halo Model through spin-dependent interaction. Assuming $m_\chi = 1\GeV$, we show the endothermic scattering of the iDM-nucleus in the left panel, with different coloured solid lines depicting different $\delta$, where  $\delta= 0\,(\mathrm{blue}),1\,(\mathrm{orange}), 2\,(\mathrm{green}),\mathrm{and}\,3\keV\,(\mathrm{red})$. The right panel shows the process of exothermic scattering, where $\delta= -2\,(\mathrm{blue}),-5\,(\mathrm{orange}), -20\,(\mathrm{green}),\mathrm{and}\,-50\keV\,(\mathrm{red})$.
}
\label{EdRdE}
\end{figure}

In Fig.~\ref{EdRdE}, we calculated the nuclear recoil spectrum of inelastic dark matter-nucleus scattering to indicate to what extent the kinematics of the iDM-nucleus affects the event rate.
For simplicity, we consider the dark matter with the mass splitting of $\delta$ and mass $m_\chi=1\GeV$ coupled to protons only.
We can see two cases of inelastic scattering with a xenon atomic target when considering the individual spin-dependent operator $\boldsymbol{O_4}$. the left panel is endothermic scattering, and the right panel is exothermic scattering. We scale the strength of the SD interaction at $c_i^p=10^{4}$ to correspond to the reference cross section $\Bar{\sigma}_{\chi p}\sim10^{-30}\cm^2$ for SD interaction at $m_\chi=1\GeV$, and the value of nuclear recoil energy $E_{R}^{max}$($E_{R}^{min}$) corresponds to $\theta=\pi\,(\theta=0)$ in Eq.~\ref{ER}. 
For endothermic scattering, based on Eq.~\ref{energy}, it can be seen that with $\delta$ increasing, $E_{R}^{max}$ decreases and $E_{R}^{min}$ increases, and the peak nuclear recoil rate is reduced accordingly. 
This indicates that kinematically elastic scattering is more favourable than endothermic scattering, and this becomes more pronounced as $\delta$ becomes larger.
For $m_\chi=1\GeV$, the maximum available initial-system kinetic energy is about $3.2\keV$, while $\delta$ is larger than this value, the rate cannot be generated, as shown in the left panel of the figure. On the other hand, in exothermic scattering with $\delta<0$, due to the fixed maximum incoming velocity of the DM, the peak of the recoil spectrum does not drop, both $E_{R}^{min}$ and $E_{R}^{max}$ increase with increasing $\left\vert\delta\right\vert$. This illustrates that for larger $|\delta|$, the scattering is more (less) kinematically favored for sufficiently small (large) energies.

\begin{figure}[ht]
    \centering
    \includegraphics[height=8cm,width=8cm]{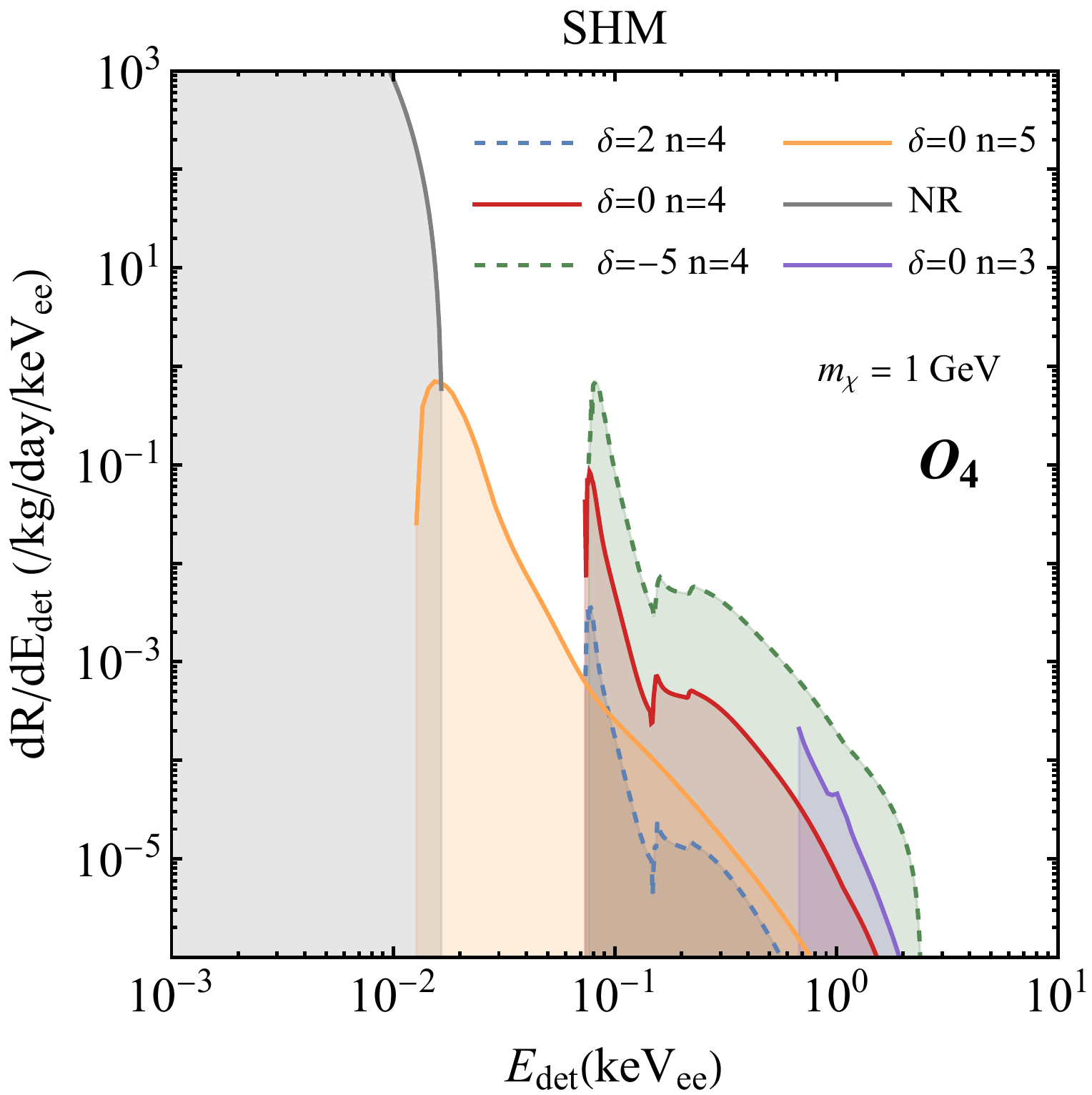}
    \caption{The rate of Migdal events induced by a mass $m_\chi=1\GeV$ dark matter particle scattering with the nucleus through the spin-dependent interaction for the xenon targets in SHM. Coloured solid lines depict the contributions of various atomic energy levels represented by $n$. For $n = 4$, we depict the endothermic (exothermic) scattering in Migdal with a blue (green) dashed line, where $\delta=2\keV$($\delta=-5\keV$). The black solid line depicts the elastic nuclear recoil.}
    \label{Migdal rate}
\end{figure}

In xenon-based detectors, ionized electrons produced by Migdal can be detected, so we depict the differential event rates as a function of the detected energy (in units of $\keV$ electron equivalent, $\keV_{\mathrm{ee}}$) by inducing the SD ($\boldsymbol{O_4}$) interaction with xenon in Fig.~\ref{Migdal rate} (consider the incoming DM particle $m_\chi=1 \GeV$). Note that to facilitate comparisons, we will scale the coupling strength to expect nuclear recoils up to $10^4$. To illustrate their characteristics, we use the solid black line represents the spectrum for nuclear recoil, the colored solid lines represent the Migdal scattering rates induced by different electron energy levels determined by $n$, and the dashed line gives the effect caused by different mass splitting $\delta$ on the same electron shell.

The Migdal scattering rate depends on the electron energy level $n$. This is because the electrons in the outer shell are more easily excited/ionized. In contrast, the electrons in the inner shell $n=3$ require higher energy to excite/ionize them, about $0.7\keV_{ee}$ as shown in Fig.~\ref{Migdal rate}.
Secondly, comparing the Migdal rates for the same electron energy level, we can see that for endothermic scattering, there is an overall decrease in the recoil spectrum and the opposite for exothermic scattering, but for both, there is no abrupt change in the shape of the spectrum. 
The peak of the Migdal rate is determined by the binding energies of the different energy levels $n$. Consequently, this quantity is determined by $n$ and is largely independent of the dark matter parameter.

Although we have included the nuclear recoil spectrum in Fig.~\ref{Migdal rate}, it is important to note that this is only to compare the Migdal spectrum ($E_{\mathrm{det}}=\mathcal{Q}E_R$ for elastic nuclear recoil). For $m_\chi=1\GeV$ dark matter, the Migdal rate becomes the dominant rate with $E_{\mathrm{det}}$ above $100\keV_{\mathrm{ee}}$, thus implying that at lower detector thresholds the Migdal effect would be more beneficial in providing an effective window for exploring low-mass dark matter.

In direct detection experiments, the number of events is closely related to the scattering cross section of dark matter. Therefore, after calculating the rate,  we will use the results of the Migdal effect and scattering after endothermic (exothermic) to give new limits on the spin EFT operator $\boldsymbol{O_4}$ for low-mass dark matter. We give the corresponding bounds based on the data provided by the XENON1T experiment. This experiment accepts two main signals: primary scintillation light (S1), which is generated by nuclear recoil and can be detected directly, and delayed proportional scintillation (S2), which is measured as a proportional signal when a drifting electron is extracted into the gas phase. The signals S1 and S2 allow for the discrimination of nuclear/electronic recoil, and electron recoils produces events with larger S2/S1 than nuclear recoils.

Here, we have used here the single ionisation channel S2 data set from XENON1T~\cite{XENON:2019gfn} for the analysis. In the S2-only case, although this reduces the background discrimination and lifetime, it allows the lower threshold to enter the analysis. This case does not distinguish between nuclear and electronic recoil, thus establishing cross section bounds for different cases of low-mass DM. Based on the experimental thresholds for XENON1T, it is reasonable to integrate the rate of events in the range $E_{\mathrm{det}}=0.186-3.8\keV_{\mathrm{ee}}$ in order to implement a single bin analysis (no signal is seen below $0.186\keV$) and to consider taking into account the energy dependent efficiency. At an exposure of 22 tonne$\cdot$day,  the expected number of events from the background was $n_{\mathrm{exp}}=23.4$, while the total number of observed events was $n_{\mathrm{obs}}=61$. From these data, using the profile likelihood ratio~\cite{Cowan:2010js} gives an upper limit of 48.9 for the number of events expected for dark matter at 90\% C.L. 
For the latest liquid xenon (LXe) detector, LUX-ZEPLIN (LZ), which has a higher sensitivity to nuclear recoil energies at the $\mathcal{O}(\keV)$ level~\cite{LZ:2019sgr,LZ:2021xov}. Therefore, we also make projections for the sensitivity of the LZ experiment assuming an exposure of $5.6\times1000$ tonne$\cdot$day. We used the energy dependent efficiency of XENON1T to integrate over the range $E_{\mathrm{det}}=0.5-4\keV_{\mathrm{ee}}$~\cite{Bell:2021zkr,LZ:2021xov,LZ:2019sgr}, an expected event rate of $2.5\times10^{-5}\mathrm{(/kg/day/keV)}$ from ${}^{220}\,\!\mathrm{Rn}$ for the background, and uncertainty of 15\% for the background~\cite{LZ:2021xov}, and finally obtain an upper limit on the expected number of events of 79.6.

\begin{figure}[t]
    \centering
    \includegraphics[height=8cm,width=7.5cm]{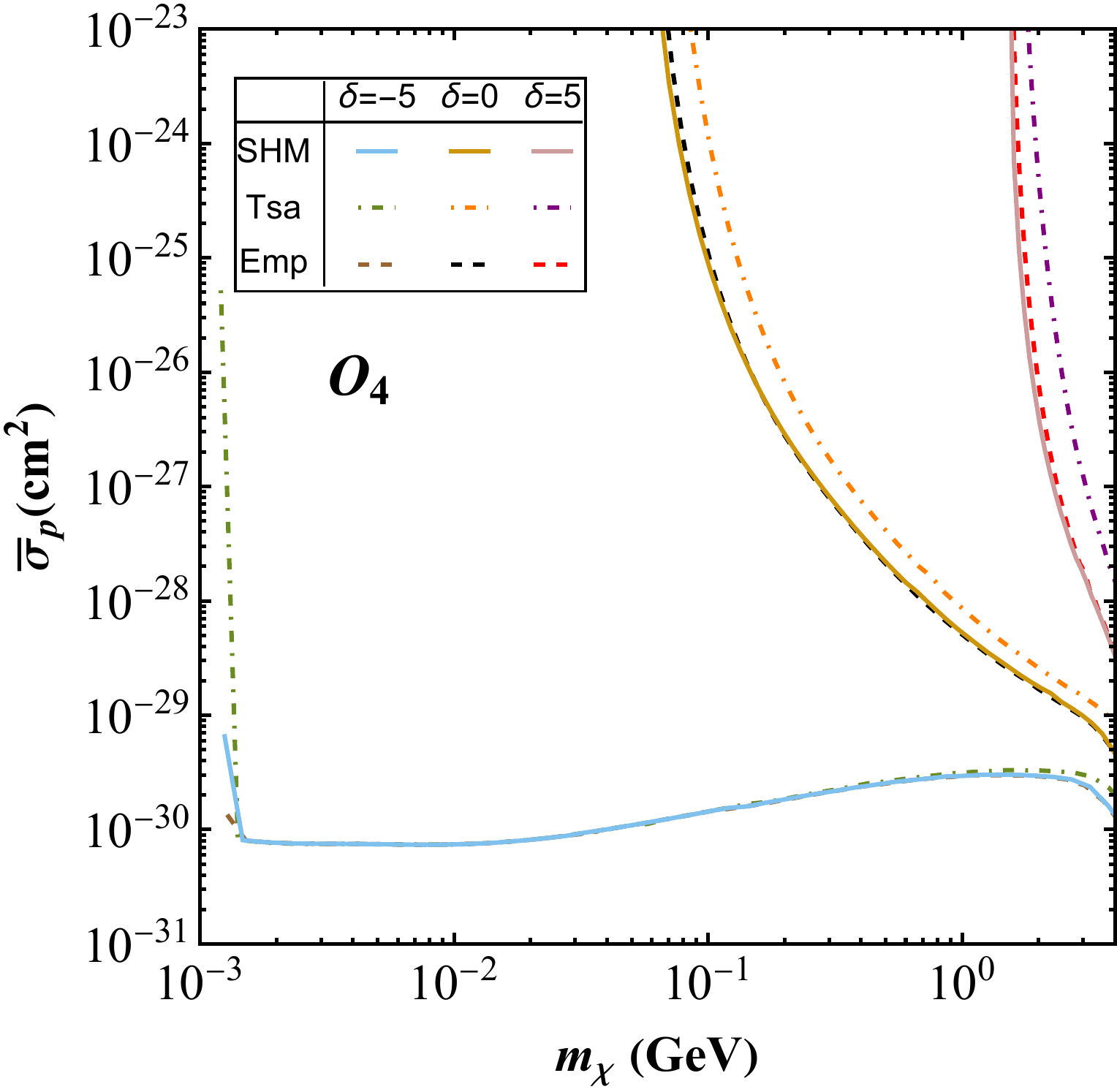}
    \hspace{1cm}
    \includegraphics[height=8cm,width=7.5cm]{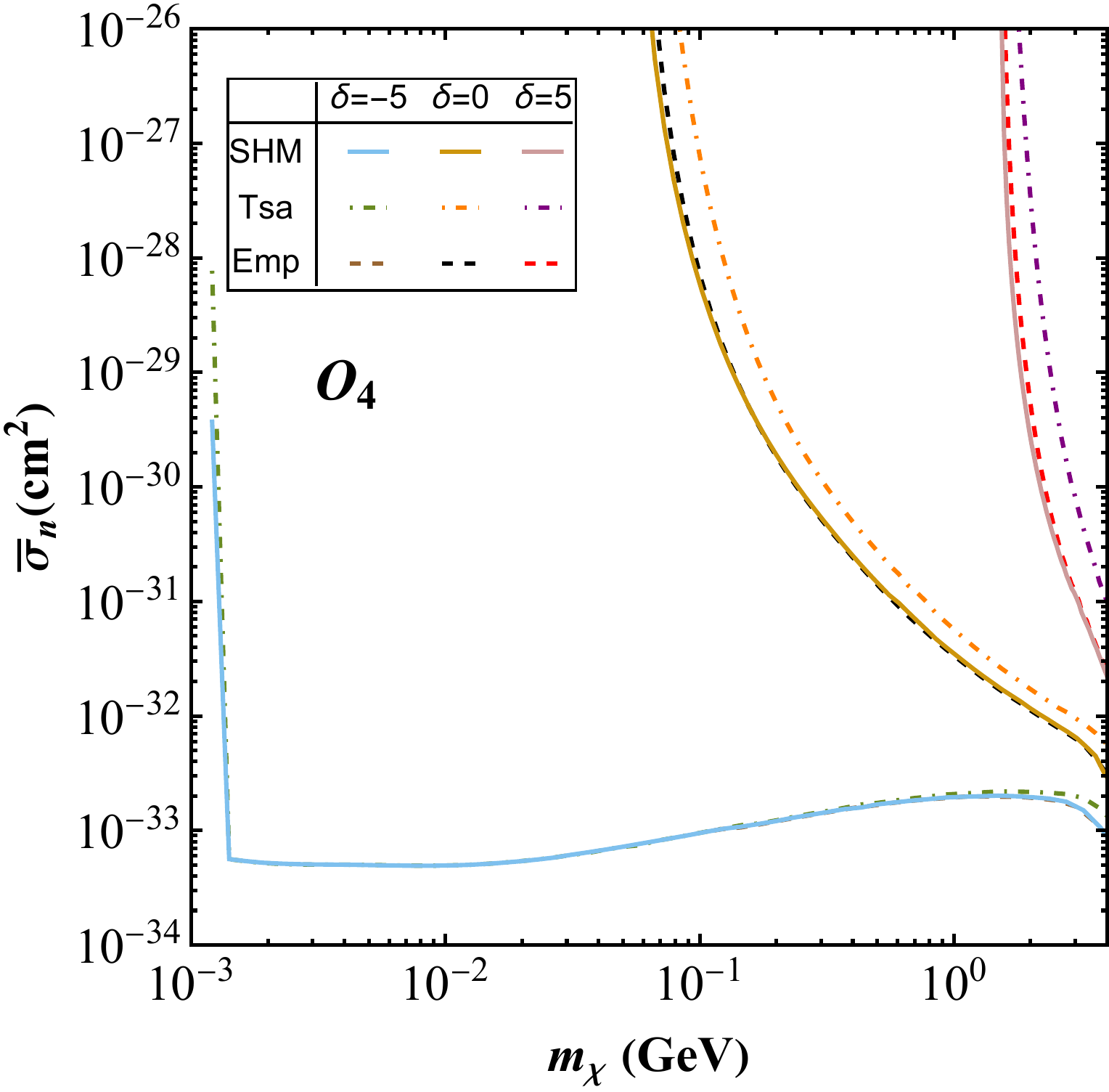}
    \caption{Constraints (90\% C.L.) from three velocity distribution functions on the iDM-nucleus Migdal effect of the spin-dependent interaction. We depict the bounds of Tsallis, Empirical and Standard Halo Model with dash-dotted, dashed and solid lines, respectively. 
    The coloured lines depict the impacts of different DM mass splitting on the bounds at each velocity distribution, where $\delta= -5, 0, 5\keV$ respectively. The left panel represents the interaction of DM with proton-only coupling, and the right panel represents the interaction with neutron-only coupling.}
    \label{VDFSIGMA}
\end{figure}
In Fig.~\ref{VDFSIGMA}, using S2-only data from XENON1T, we compare the effects of three velocity distribution models on the iDM-nucleus Migdal scattering cross section in the spin-dependent interactions.
We have marked the SHM with a solid line, the Empirical model with dashed line and the  Tsallis model with dash-dotted line. We can observe that, as an overall trend, the bounds of the Tsallis model is weaker than that of the SHM and Empirical model for all three interactions: endothermic ($\delta=5\keV$), elastic ($\delta=0\keV$), and exothermic ($\delta=-5\keV$), with this difference more apparent in the elastic and endothermic. There is even an order of magnitude difference between them. The empirical model is only slightly stronger than the SHM limit above the DM mass of $0.1\GeV$ (for $\delta=0\keV$). 
These situation can be traced back to Fig.~\ref{VDF}, where the $\eta(v_{\rm{min}})$ of the Tsallis model falls more rapidly at greater than $300\,\rm{km/s}$.
Therefore, for larger $v_{\rm{min}}$ (see Eq.~\ref{Vmin}), the smaller the value of $\eta(v_{\rm{min}})$, the weaker the associated generating bounds.
Returning to Eq.~\ref{Vmin}, exothermic scattering makes it easier for DM with masses in the $10^{-3}\sim1\GeV$ region to fall in the low-velocity region.
The three models almost overlap for DM masses below $1 \GeV$, indicating that lower masses of DM retain more flexibility in the choice of VDFs. Compared with the spin-independent results in Ref.~\cite{Bell:2021zkr}, we found that our spin-dependent Migdal scattering cross section is much weaker but has a similar slope.
\begin{figure}[ht]
    \centering
    \includegraphics[height=8cm,width=7.5cm]{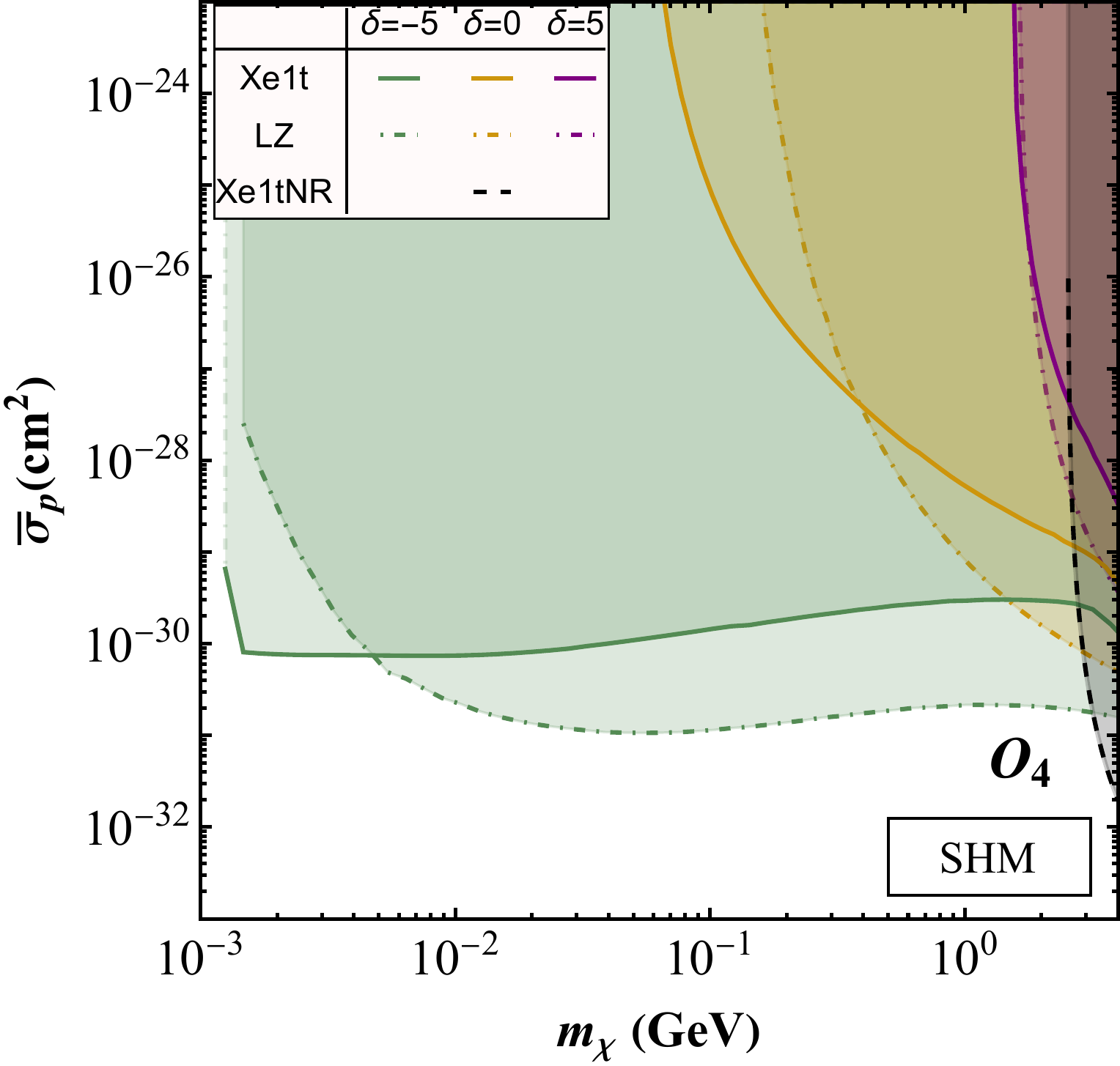}
    \hspace{1cm}
    \includegraphics[height=8cm,width=7.5cm]{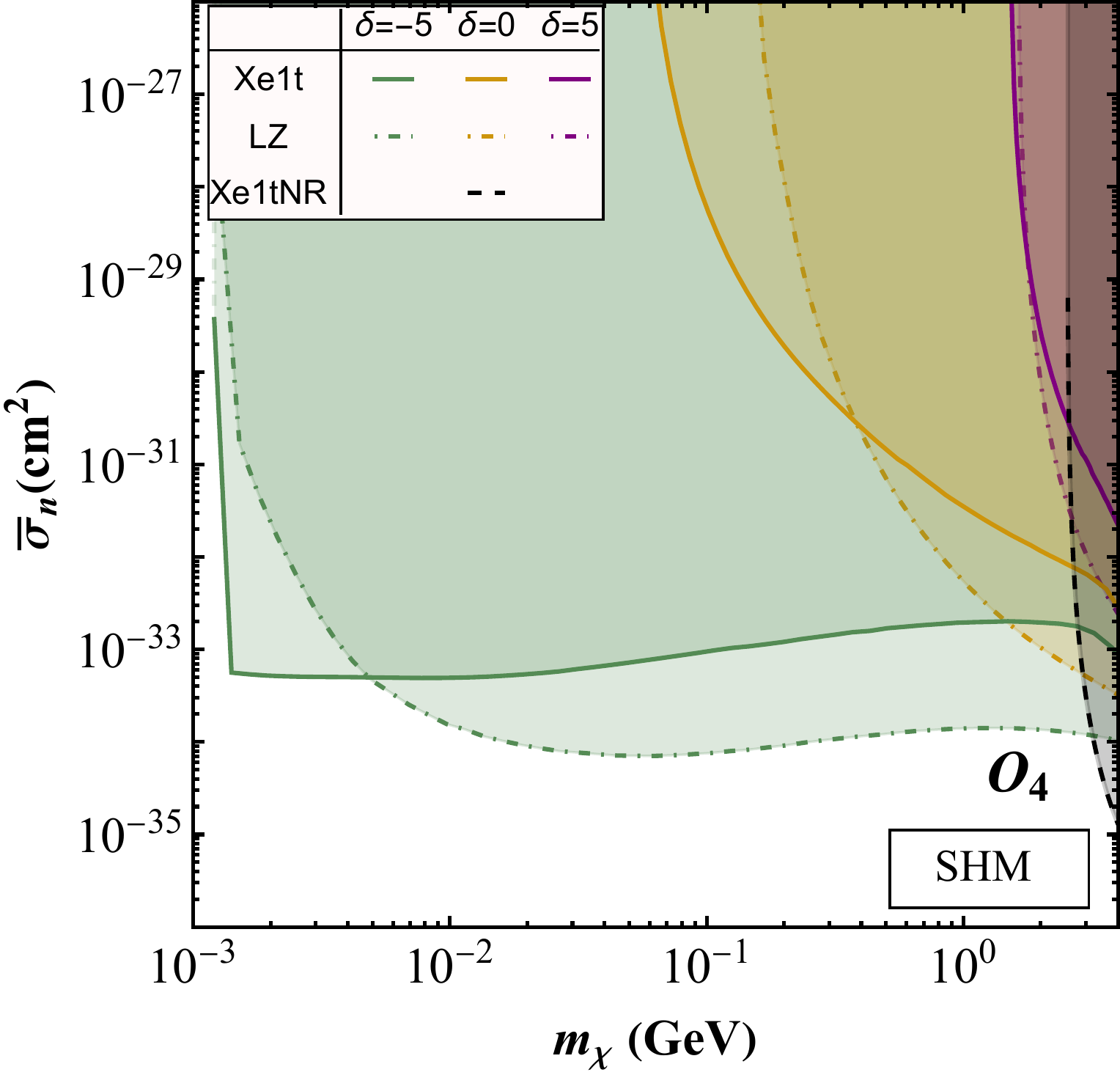}
    \caption{The 90\% C.L. limits on the iDM-nucleus scattering cross section of spin-dependent interaction from XENON1T on nuclear recoils (dashed), XENON1T with the Migdal effect (solid) and the projected LZ sensitivity with the Migdal effect (dash-dotted). The coloured lines depict the different scattering processes, where  $\delta=5\keV$ for endothermic (purple), $\delta=-5\keV$ for exothermic (green), and $\delta=0\keV$ for elastic scattering (orange). We plot the scattering cross sections of the iDM coupling with protons and neutrons in the left and right panels, respectively.}
    \label{sigma}
\end{figure}
    
In Fig.~\ref{sigma}, for the spin-dependent operator $\boldsymbol{O_4}$, after accounting for elastic, exothermic and endothermic interactions, we show the constraints on the cross section for XENON1T and LZ experiments coupled with protons and neutrons alone at 90\% C.L., respectively. Even for different couplings, the various bound shapes are quite similar, and the elastic scattering provides a good analogy: the Migdal effect shows more limits than the elastic nuclear recoil around $m_\chi\simeq3\GeV$, below which the bounds all originate from the Migdal effect and yield more bounds for the lower masses of dark matter. In fact, the crossover point between the Migdal and nuclear recoil boundary is determined by the threshold reached by the detector. Under the spin-dependent operator $\boldsymbol{O_4}$, the degree of constraint on the proton and neutron cross sections differs, with the cross section of the proton being weaker than that of the neutron since the xenon isotope, which has an even number of protons $(Z = 54)$, the spin expectation value of the proton is smaller after the intrinsic spin magnetic moments offset each other.

Our work discusses the SD interactions of the $\boldsymbol{O_4}$ operator and uses the nuclear form factor of ${}^{131}\!\rm{Xe}$ from Ref.~\cite{Anand:2013yka}. In the formalism of Haxton et al., they encode the nuclear physics part into the nuclear response function. It is worth noting that our used nuclear form factor includes the effect of the one-body current, and a truncation of the valence space is made in the calculation. Recently, in Ref.~\cite{Klos:2013rwa}, Klos et al. applied large-scale shell-model calculations to evaluate the nuclear form factor for DM-nucleon SD interactions with chiral one- and two-body currents (1BCs and 2BCs). And B. S. Hu et al.~\cite{Hu:2021awl} used the valence-space formulation of the in-medium similarity renormalization group to calculate ab initio spin-dependent form factors for all nuclei currently used in the direct detection searches. For ${}^{131}\!\rm{Xe}$, the form factors obtained for Ref.~\cite{Hu:2021awl} are consistent with the results in Ref.~\cite{Klos:2013rwa} at the 2BCs level. By comparison with Ref.~\cite{Klos:2013rwa}, we find that the ${}^{131}\!\rm{Xe}$ neutron form factors we use are slightly larger than them at momentum $q\lesssim100 \rm{MeV}$, but the overall curve difference is not significant. However, there is a large difference in the form factor of protons, and the chiral 2BCs lead to a rather significant enhancement effect. With the form factor in Ref.~\cite{Hu:2021awl}, the LZ collaboration reported their exclusion limits for the SD interacions~\cite{LZ:2022ufs}. Since we used the nuclear form factor in~\cite{Anand:2013yka}, the difference between our ``proton-only" and ``neutron-only" results is nearly $10^3$, instead of 30 in Ref.~\cite{LZ:2022ufs}.

On the one side, we notice that the bound of endothermic scattering is very closer to that of elastic scattering. According to Eq.~\ref{maxEem}, as the mass loss $\delta$ increases in endothermic scattering, the approximate maximum available energy projected to the Migdal electron decreases, so the bounds loses sensitivity to low-mass dark matter more rapidly.
On the other side, Migdal electrons in exothermic scattering can acquire more energy so that they still have above-threshold sensitivity at lower DM masses. Essentially, Migdal electrons are easily excited above the threshold because for DM with masses below GeV, $\frac{1}{2}m_\chi v^2\sim\mathcal{O}(\keV)$, then $\delta\sim\mathcal{O}(\keV)$, there is a significant enhancement effect. 
For DM masses below $m_\chi\sim 7\MeV$, the limiting boundary of LZ is weaker than that of XENON1T, since we predict a higher threshold for LZ $(E_{\mathrm{det}}\ge0.5\keV_{\mathrm{ee}})$ than for the S2-only analysis of XENON1T. We also clarify that an S2-only analysis of the LZ experiment could improve the sensitivity to lighter dark matter masses. However, we need to know more about the achievable thresholds, backgrounds, and exposures for this S2-only analysis.

In addition, we set a cut-off value for the $E_R$ in Migdal process, as mentioned in Ref.~\cite{Knapen:2020aky}.
In calculating the ionisation function, for the impulse approximation to hold, it is necessary to ensure that the collision time $t_{\mathrm{collsion}}\sim 1/E_R$ is less than the time $t_{\mathrm{traverse}}\sim 1/\omega_{ph}$ ($\omega_{ph} $ is the phonon frequency) for the atom to traverse its potential field.
For sufficiently small DM masses, there will be recoil energy falling below this cut-off value and the Migdal rate will fail, so this value has a relatively large impact on the detector threshold as well as on the low dark matter masses.
Referring to the method in Ref.~\cite{Bell:2021zkr}, use the time $t\sim\mathcal{O}(10^{-12})s$ required for a xenon atom to traverse the average interatomic distance at 170k at the speed of sound as the cut-off time. We conservatively to set $E_{R\,\mathrm{cut}}\ge50\,\mathrm{meV}$.
Thus, we can place a limit on the mass of dark matter: elastic scattering corresponds to 0.02 GeV, while exothermic (endothermic) scattering relies on the mass splitting $\delta=-5\keV\,(5\keV)$, which is $0.001\GeV\,(0.36\GeV)$.

\section{Inelastic Dark Matter-Electron Scattering}
\label{idme}
This section will investigate inelastic dark matter-electron scattering~\cite{Baryakhtar:2020rwy,Bramante:2020zos,Chao:2020yro,Harigaya:2020ckz,Dror:2020czw} in a non-relativistic effective field theory. We will briefly discuss the relevant kinematics and derive the formulae for our calculations.

\subsection{Calculations}
According to our previous description, inelastic dark matter-electron scattering is very similar to the previously described inelastic dark matter-nucleus scattering process. The electron spectrum of Migdal is evaluated in terms of the effective transfer momentum $\vec{q}_e\equiv\frac{m_e}{m_N}\vec{q}$, which is the most significant difference from electron scattering. For the energy conservation of the iDM-electron scattering process, it is simple to rewrite Eq.~\ref{N} as $ E_{\rm{em}}+\delta=\vec{q}\cdot \vec{v}  -\frac{q^2}{2\mu_{\chi e}}$, where $\mu_N\to\mu_{\chi e}$ ($\mu_{\chi e}$is the reduced mass of the DM and
electron).
Furthermore, when the maximum incoming velocity $v_{\rm{max}}$ of the DM is fixed, we can determine the range of allowed momentum transfers. The minimum and maximum momentum transfer are
\begin{equation}
\begin{split}
q_{\mathrm{min} }&=\mathrm{sign}\left ( E_{\mathrm{em}}+\delta  \right )\,m_\chi v_{\mathrm{max} }\left ( 1-\sqrt{1-\frac{2\left ( E_{\mathrm{em}}+\delta  \right )}{m_\chi v^2_{\mathrm{max} }} }  \right )\\\\
q_{\mathrm{max} }&=m_\chi v_{\mathrm{max} }\left ( 1+\sqrt{1-\frac{2\left ( E_{\mathrm{em}}+\delta  \right )}{m_\chi v^2_{\mathrm{max} }} }  \right ).
\end{split}
\label{Eqlimit}
\end{equation}
In the limit $\delta\to0$ Eq.~\ref{Eqlimit} reduces to elastic scattering.

Similarly to nuclear, we also introduce an effective field theory for iDM-electron scattering according to the work of Catena et al~\cite{Catena:2019gfa}. In this formalism, the active degrees of freedom will be DM particles and electrons.
The symmetry governing non-relativistic DM-electron scattering is replaced by Galilean invariance instead of Lorentz invariance under relativistic boosted.
Thus, the invariant amplitude of DM-electron scattering can still be represented by a series of operators consisting of Galilean invariants. 

In this EFT, there are also four three-momentum Galilean invariants: $\vec{q}$, $\mathbb{\vec{S}}_e$, $\mathbb{\vec{S}}_e$, $\vec{v}^{\perp}_{\mathrm{inel}}$.
Here corresponding to the inelastic case, $\vec{v}^{\perp}_{\mathrm{inel}}$ is defined as
\begin{equation}
   \vec{v}^{\perp}_{\rm{inel}}\equiv\vec{v}+\frac{\vec{q}}{2\mu_{\chi e}}+\frac{\Delta}{|q|^2}\,\vec{q}
\label{vel}
\end{equation}
Since the conservation of energy in the iDM-electron scattering process, $\vec{v}^{\perp}_{\rm{inel}}\cdot\vec{q}=0$.
Compared with the definition of Eq.~\ref{Ninel}, we can see that the process of inelastic scattering of electrons only modifies $\mu_N\to\mu_{\chi e}$. Also, this modification is only reflected in the DM particle response function $\mathscr{R}^{nl}_i$. For the operator $\boldsymbol{O}_4=\mathbb{\vec{S}}_e\cdot\mathbb{\vec{S}}_\chi$, it is not subject to $\vec{v}^{\perp}_{\rm{inel}}$, so we can still refer to the results of the elasticity calculation in Ref.\cite{Catena:2019gfa}.

It is worth noting that in this EFT, the invariant scattering amplitude $\mathcal{M}(\vec{q},\vec{v}^\perp_{\rm{inel}})$ of the DM-electron does not depend explicitly on the characteristics of the specific mediator particle. However, this formalism is still applicable when the mediator particle mass is much larger than the transfer momentum: $m^2_{\rm{med}}\gg q^2$ (contact interaction), or much smaller than the transfer momentum: $m^2_{\rm{med}}\ll q^2$ (long-range interaction)~\cite{Li:2014vza}.
To summarise these, the free amplitudes of non-relativistic iDM-electron scattering express as
\begin{equation}
    \mathcal{M}(\vec{q},\vec{v}^\perp_{\rm{inel}})=\sum_i\left(c^s_i+c^l_i\,\frac{q^2_{\mathrm{ref}}}{|q|^2}\right)\langle \boldsymbol{O}_i\rangle,
\label{two interaction}
\end{equation}
where the reference momentum $q_{\mathrm{ref}}\equiv \alpha m_e$, $\alpha=1/137$, and the coefficient $c^s_i$ ($c^l_i$) represents the contact (long-range) interaction of the DM particle with the electron.

To obtain the total event rate we are interested in, first, the total transition rate of electrons induced by DM for the initial state of electron $|e_1\rangle$ $\to$ final state $|e_2\rangle$ is
\begin{equation}
    \mathbb{R}_{1\to2}=\frac{n_\chi}{16m^2_\chi m^2_e}\int\frac{\mathrm{d}^3q}{(2\pi)^3}\int \mathrm{d}^3v\,f(v)\,(2\pi)\,\delta(E_f-E_i)\,\overline{|\mathcal{M}_{1\to2}|^2},
\label{R1-2}
\end{equation}
where $n_\chi = \rho_\chi/m_\chi$ is the local DM number density, $E_f \,(E_i)$ is the final (initial) state energy of the system, and the $\delta$ function ensures the conservation of energy for this process.
$\overline{|\mathcal{M}_{1\to2}|^2}$ was defined as the squared electron transition amplitude~\cite{Catena:2019gfa},
\begin{equation}
    \overline{|\mathcal{M}_{1\to2}|^2}=\overline{\left\vert\int\frac{\mathrm{d}^3k}{(2\pi)^3}\,\psi^*_2(\vec{k}+\vec{q})\,\mathcal{M}(\vec{q},\vec{v}^\perp_{\rm{inel}})\,\psi_1(\vec{k})\right\vert^2}.
\end{equation}
Here $\psi_1$ and $\psi_2$ represent the electron initial and final state wave functions, respectively, and this equation has been averaged (summed) over the initial (final) spin states. Then, we can write down the iDM-electron scattering differential event rates that include the full atomic orbitals
\begin{equation}
\begin{split}
    \frac{\mathrm{d}R}{\mathrm{d}\ln E_e}&=N_T\sum^l_{m=-l}\sum^\infty_{l^\prime=0}\sum^{l^\prime}_{m^\prime=-l^\prime}\frac{Vk^{\prime\,3}}{(2\pi)^3}\mathbb{R}_{1\to2}\\
    &=N_T\,\frac{n_\chi}{128\pi\,m^2_\chi m^2_e}\int \mathrm{d}q\,q\int \frac{\mathrm{d}^3v}{v}\,f(v)\Theta(v-v_{\rm{min}})\overline{|\mathcal{M}^{nl}_{\rm{ion}}|^2},
\end{split}
\label{ionspectrum}
\end{equation}
where $N_T$ is the number of target atoms, $V=(2\pi)^3\delta^3(0)$ is the normalized phase space factor~\cite{Essig:2015cda} and $\Theta$ is a step function to ensure that the incoming speed of the DM reaches the energy required to cause the electron recoil. And $\overline{|\mathcal{M}^{nl}_{\rm{ion}}|^2}$ is the so-called electron ionisation amplitude squared, defined as
\begin{equation}
\begin{split}
    \overline{|\mathcal{M}^{nl}_{\rm{ion}}|^2}&\equiv V\frac{4k^{\prime\,3}}{(2\pi)^3}\sum^l_{m=-l}\sum^\infty_{l^\prime=0}\sum^{l^\prime}_{m^\prime=-l^\prime}\overline{|\mathcal{M}_{1\to2}|^2}\\
    &=\sum^4_{i=1}\mathscr{R}^{\,nl}_i\left(\vec{v}^\perp_{\rm{inel}},\frac{\vec{q}}{m_e}\right)\mathscr{W}^{nl}_i\left(k^\prime,\vec{q}\right),
\end{split}
\label{Mion}
\end{equation}
where getting from the first expression to the second is actually a Taylor expansion of $\mathcal{M}(\vec{q},\vec{v}^{\perp}_{\rm{inel}})$ at $\vec{k}=0$, which is then expressed as a product of the DM particle response function $\mathscr{R}^{nl}_i$ and the associated atomic response function $\mathscr{W}^{nl}_i$. This approach allows for a more intuitive examination of the DM-electron scattering process.

In fact, there are four atomic response functions that can be derived from Ref.\cite{Catena:2019gfa}. In our work, only $\mathscr{W}^{nl}_{1}$ was applied
\begin{equation}
\begin{split}
    \mathscr{W}^{nl}_1(k^\prime,\vec{q})&\equiv V\frac{4k^{\prime\,3}}{(2\pi)^3}\sum^l_{m=-l}\sum^\infty_{l^\prime=0}\sum^{l^\prime}_{m^\prime=-l^\prime}|f_{1\to2}(q)|^2.\\
\end{split}
\end{equation}
For $\mathscr{W}^{nl}_1$, it is actually the ionization factor commonly used in various light dark matter detection literatures.
The $f_{1\to2}\left(\vec{q}\right)$ in the above expression is called the scalar form factor
\begin{equation}
\begin{split}
    f_{1\to2}\left(\vec{q}\right)&=\int\frac{\mathrm{d}^3k}{(2\pi)^3}\psi^*_2 (\vec{k}+\vec{q})\psi_1(\vec{k}).\\
\end{split}
\end{equation}
Corresponding to our calculation, the DM response function is $\mathscr{R}^{nl}_1\equiv\frac{j_\chi(j_\chi+1)}{12}\cdot3c^2_4$.

\subsection{Numerical Results and Discussions}

The non-relativistic effective theory of iDM-electron interactions described in the previous subsection culminates in a general expression for the electron ionization energy spectrum of isolated atoms constructed from Eq.~\ref{ionspectrum}. This almost model-independent framework and a general expression for the scattering amplitude consisting of a series of effective operators in Eq.~\ref{Mion} allow us to make predictions for direct searches for sub-GeV DM particles. For comparison purposes, we keep to the formalism in Ref.~\cite{Catena:2019gfa} and also give a reference cross-section for the electron,
\begin{equation}
    \Bar{\sigma}_e\equiv\frac{\mu^2_{\chi e}c^2_i}{16\pi m^2_\chi m^2_e}.
\end{equation}
This definition differs from the reference cross section of the nucleus, where $c_i$ does not require additional compensation for the dimensions.
The contact and long-rang interaction can then be identified using individual EFT operators and the connection between the EFT coefficients in Eq.~\ref{two interaction}. In particular, we take into account the effects of inelasticity to compare the electron ionization events induced within the detector threshold.
\begin{figure}[ht]
    \centering
    \includegraphics[height=7.5cm,width=7.5cm]{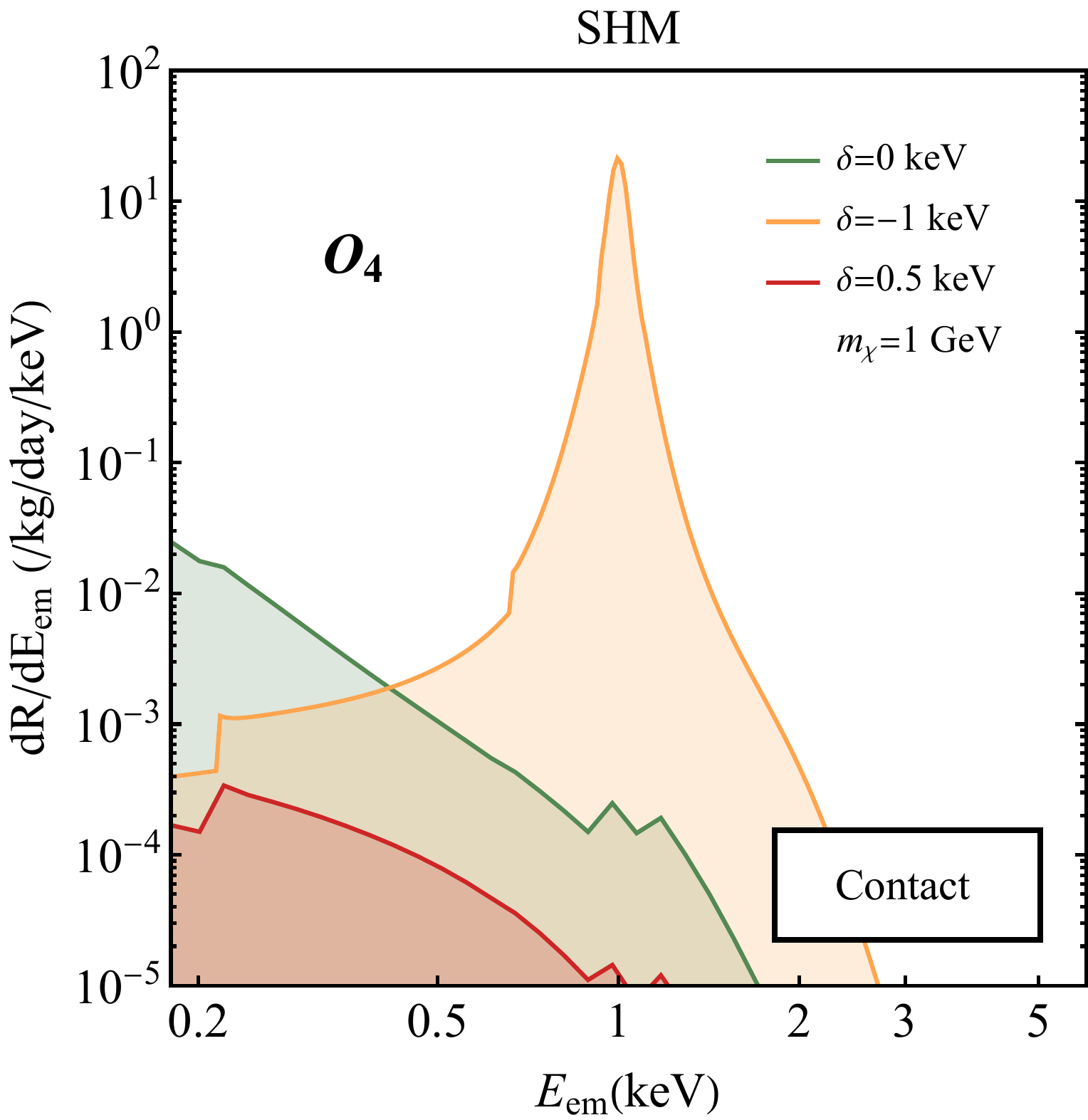}
    \hspace{1cm}
    \includegraphics[height=7.5cm,width=7.5cm]{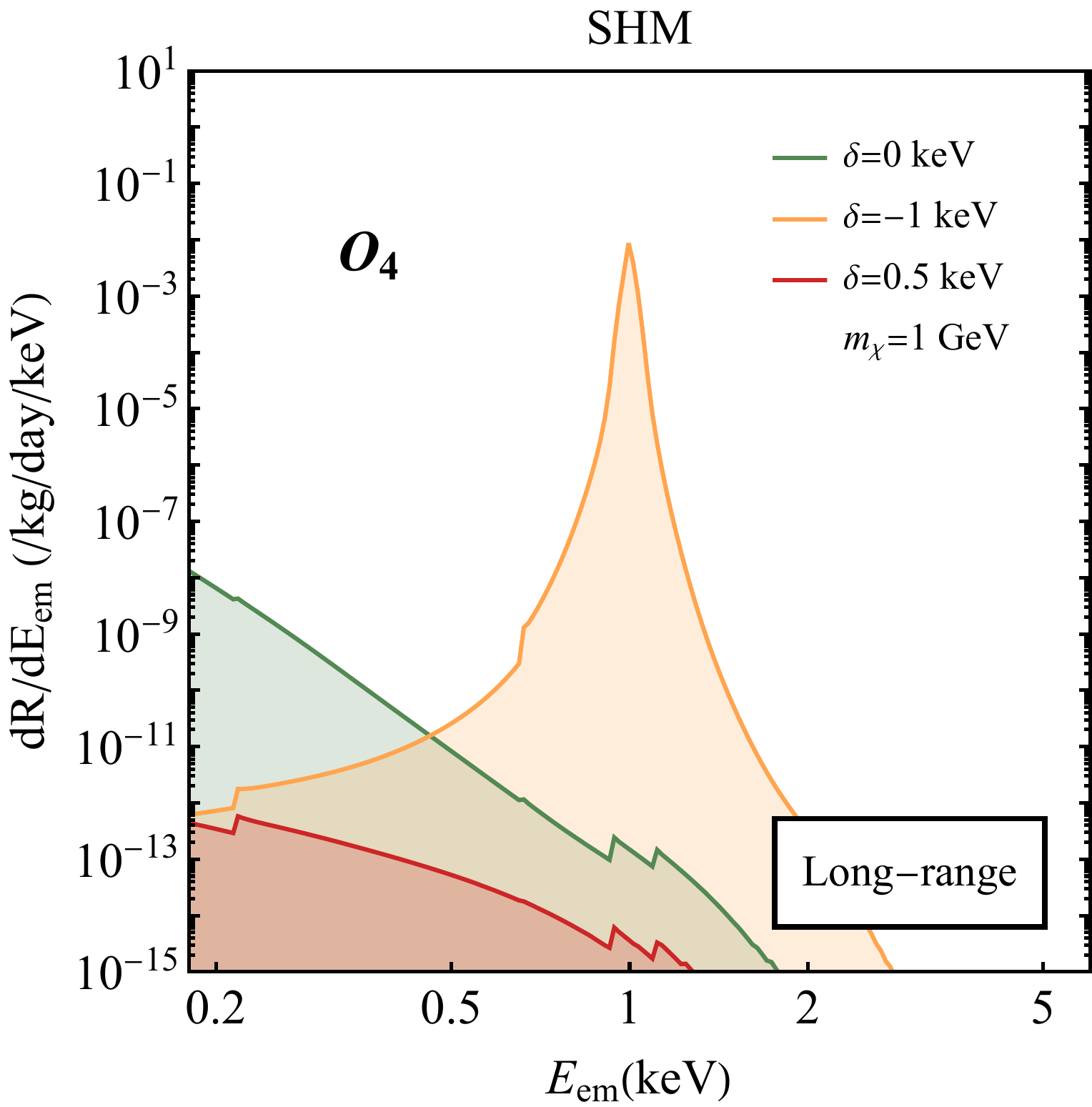}
    \includegraphics[height=7.5cm,width=7.5cm]{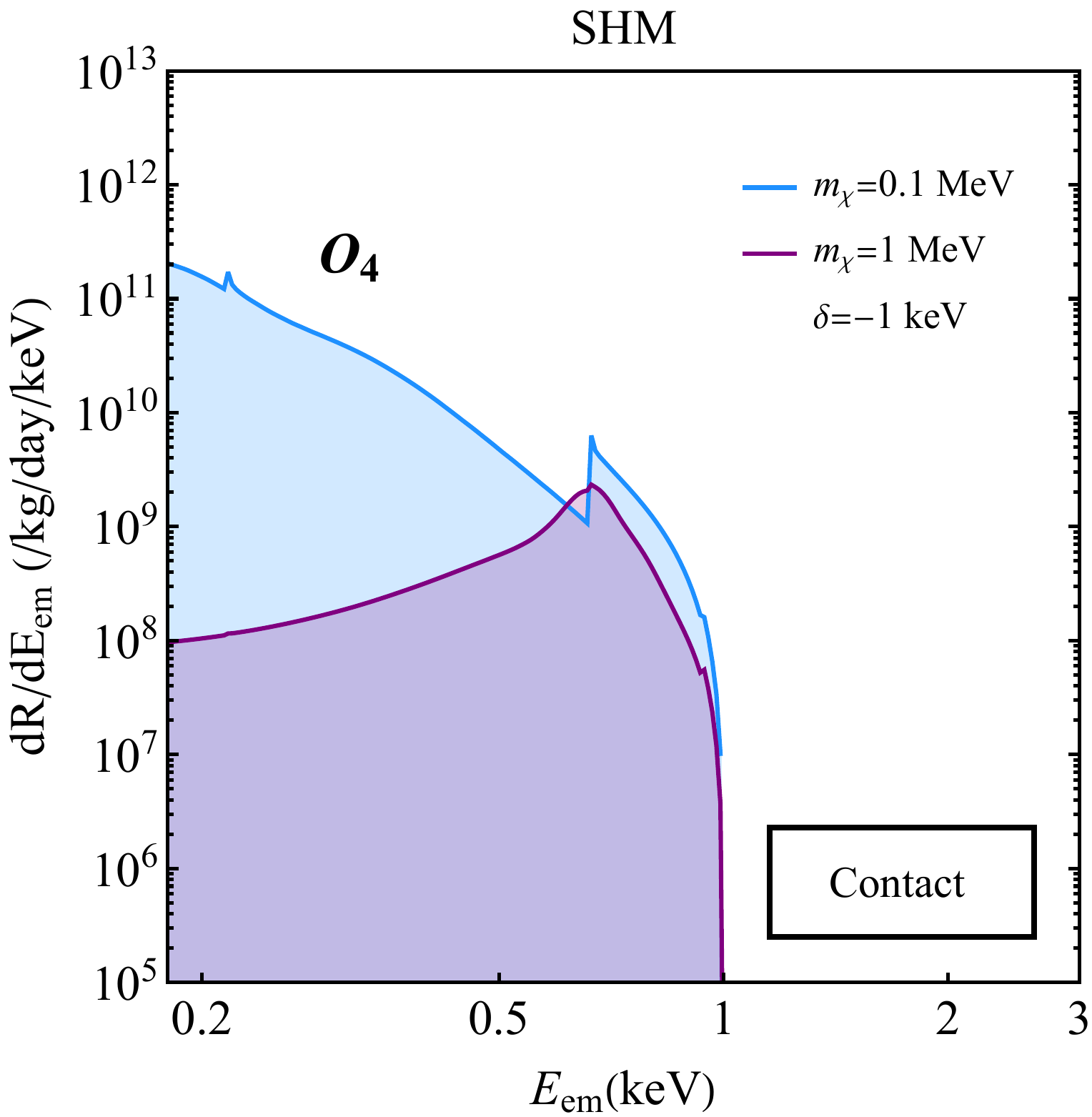}
    \hspace{1cm}
    \includegraphics[height=7.5cm,width=7.5cm]{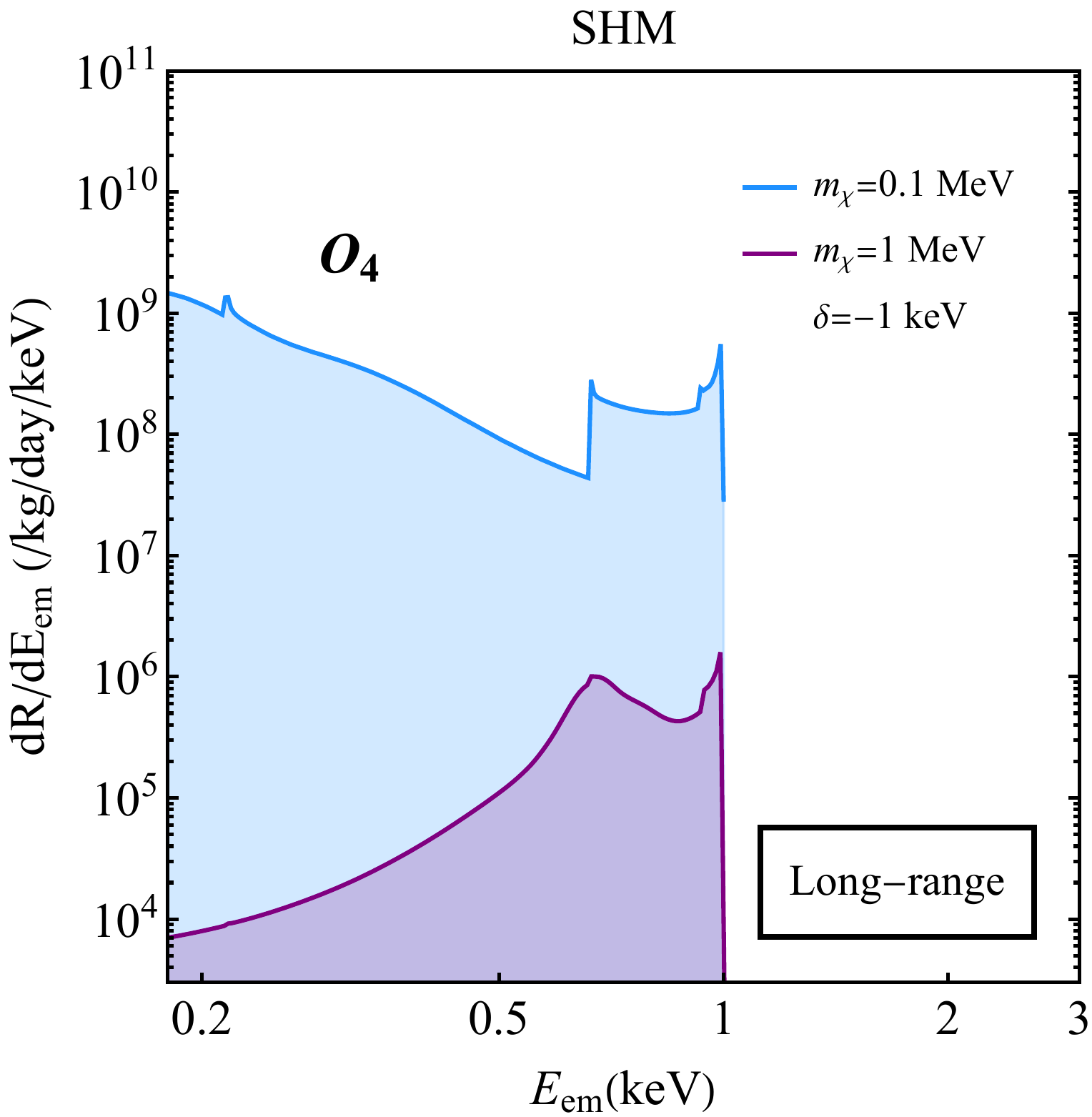}
    \caption{The differential event rate of electron scattering between DM with different masses under spin-dependent operator $\boldsymbol{O_4}$ via contact (left) and long-range (right) interaction, respectively. In the top panel, the endothermic (red), elastic (green), and exothermic (orange) scattering are depicted with solid lines of various colors corresponding to $\delta=0.5,\,0,\,-1\keV$, respectively. The bottom panel shows the differential event rate of exothermic ($\delta=-1\keV$) scattering for DM mass $m_\chi=0.1\MeV$ (blue) and $m_\chi=1\MeV$ (purple).}
    \label{dRdEe}
\end{figure}

In Fig.~\ref{dRdEe}, we used $\delta=0,-1,0.5\keV$ as fiducial parameters to show the differential event rates of exothermic, elastic, and endothermic scattering for different masses of DM with electrons through contact and long-range interactions under an individual operator $\boldsymbol{O_4}$. Here we set the coefficient $c_4=10^{-5}$ for $\boldsymbol{O_4}$, corresponding to the spin cross section $\Bar{\sigma}^{\mathrm{SD}}_e\sim\mathcal{O}(10^{-40})\,\mathrm{cm}^2$, and assume that the DM particles obey the SHM velocity distribution. In the bottom panel of Fig.~\ref{dRdEe}, only the results of exothermic scattering are shown because the DM with mass $m_\chi\leq1\MeV$ cannot produce enough recoil energy to obtain detectable electrons for elastic and endothermic scattering. 
In exothermic scattering, the event spectrum at the top panel of Fig.~\ref{dRdEe} shows a sharp peak at $E_{\rm{em}}=|\delta|$. This relationship can be understood from Eq.~\ref{Eqlimit}: when DM with mass $m_\chi=1\GeV$ can produce enough electron recoil energy,$E_{\mathrm{em}}=|\delta|$, the lower limit of transfer momentum $q_{\mathrm{min}}=0$ and the upper limit $q_{\mathrm{max}}=2m_\chi v_\mathrm{max}$, leading to the maximum integration interval. This results in a significant enhancement of the scattering rate due to the ionization function's integration over $\vec{q}$. Note that this enhancement is a feature of exothermic scattering. Furthermore, in iDM-electron scattering, the ionization event rate is severely suppressed for endothermic compared to elastic scattering and is more significant for long-range interactions. This is because for elastic and endothermic scattering, they produce typical recoil energies $\mu_{\chi e} v^2_{max}\sim\mathcal{O} (\mathrm{eV})$. Endothermic scattering does not have better sensitivity than elastic for usual Xenon-type detectors. 
It should be noted that the factor of 3 in the DM response function $\mathscr{R}^{nl}_1\equiv\frac{j_\chi(j_\chi+1)}{12}\cdot3c^2_4$ for the operator $\boldsymbol{O_4}$ is based on the assumption of non-relativistic and independent particle approximations~\cite{Catena:2019gfa}. Including the many-body effects and relativistic corrections, such a factor will be mildly changed with the variation of the electron energy~\cite{Liu:2021avx}. For simplicity, we use a constant factor of 3 in the calculations.

\begin{figure}[t!]
    \centering
    \includegraphics[height=6.5cm,width=5.4cm]{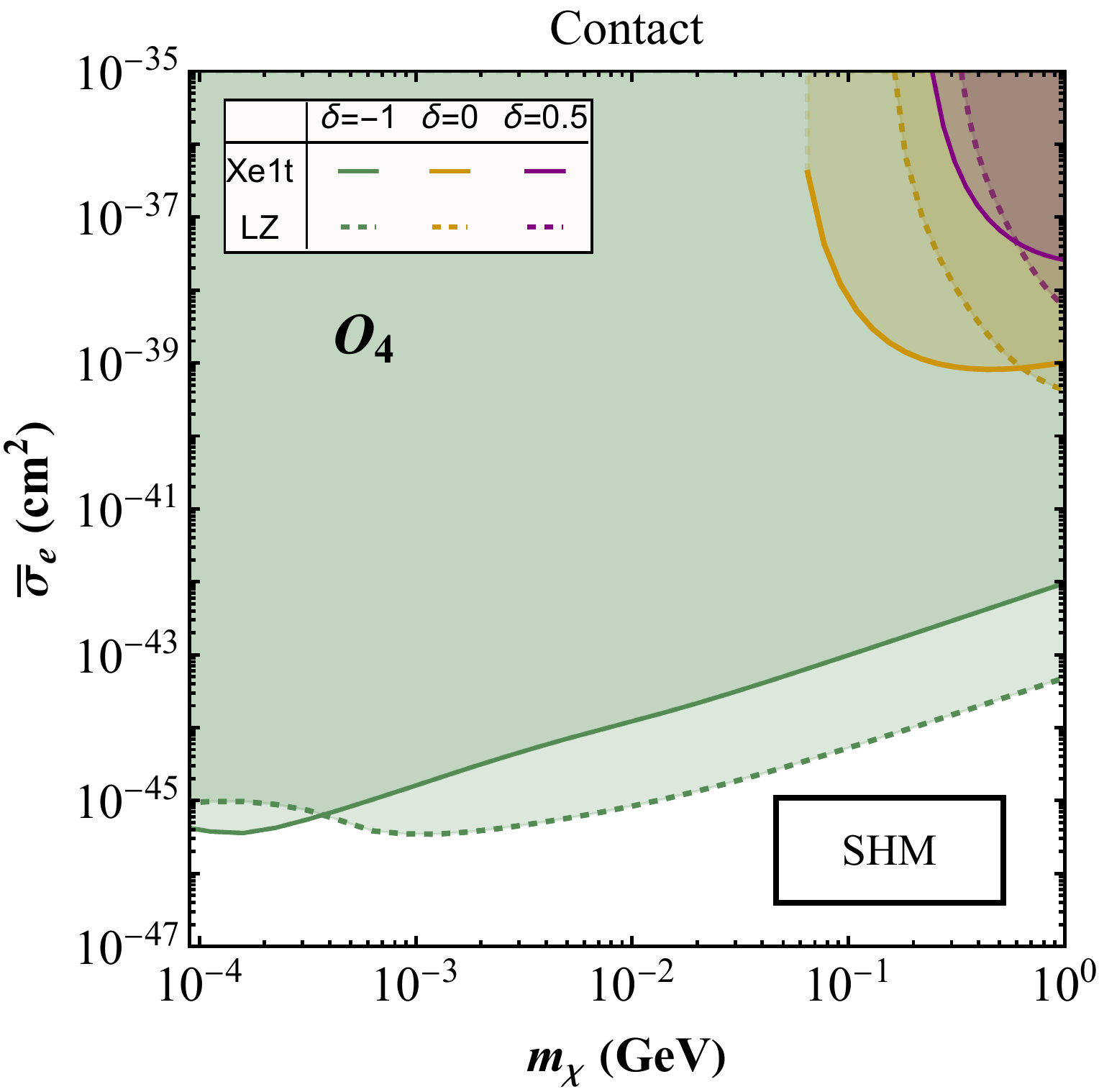}
    \includegraphics[height=6.5cm,width=5.4cm]{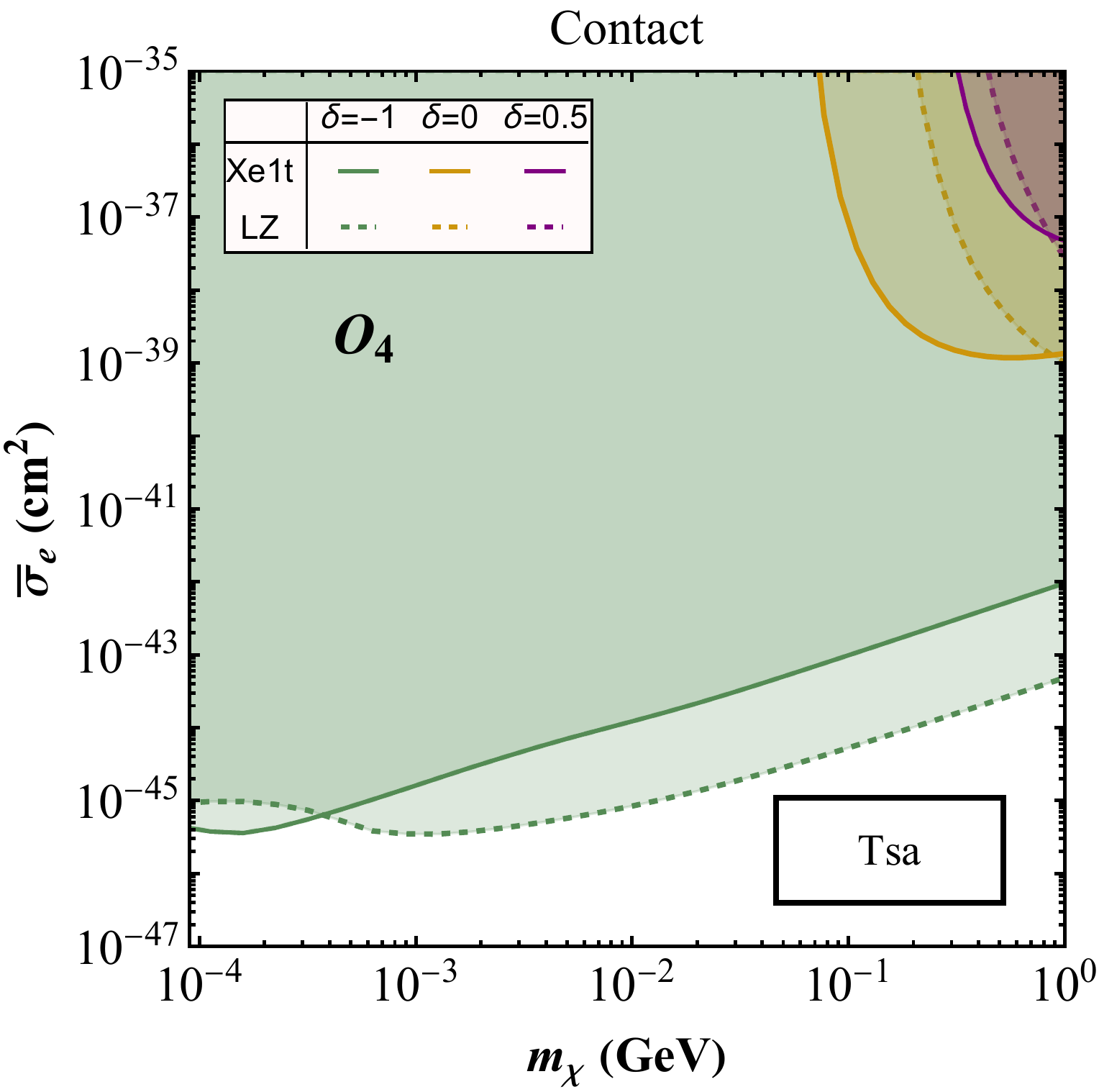}
    \includegraphics[height=6.5cm,width=5.4cm]{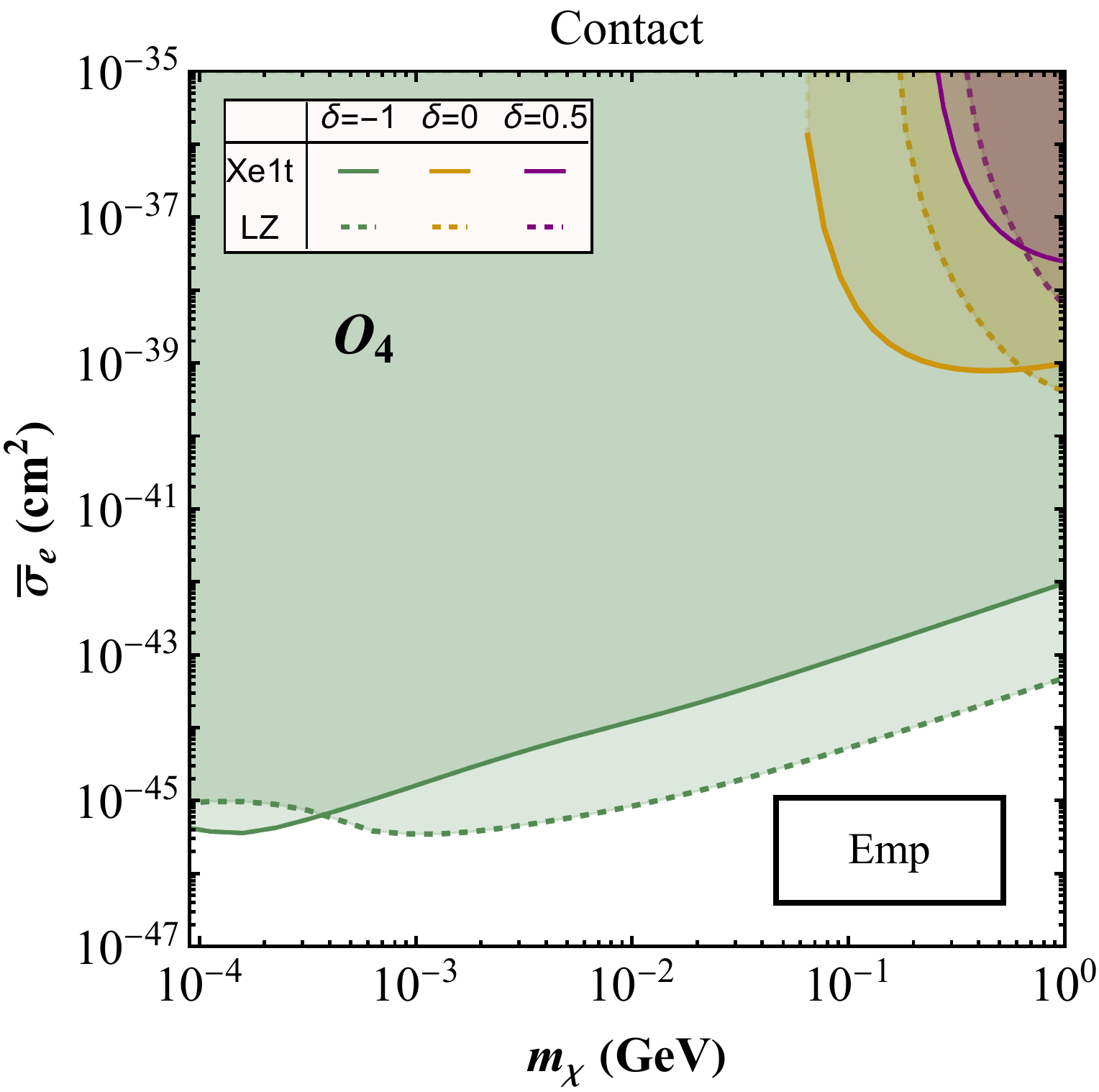}
    \includegraphics[height=6.5cm,width=5.4cm]{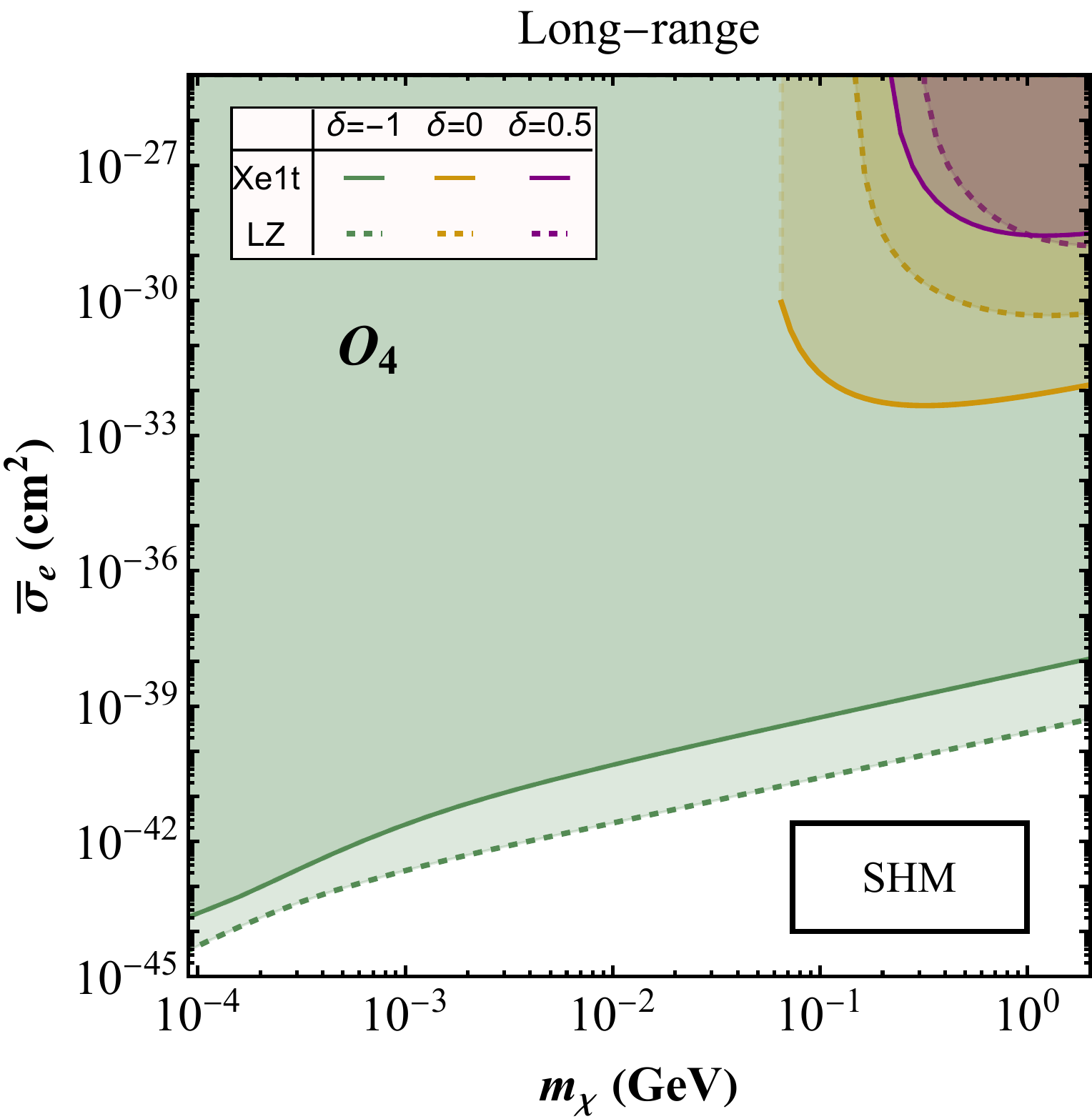}
    \includegraphics[height=6.5cm,width=5.4cm]{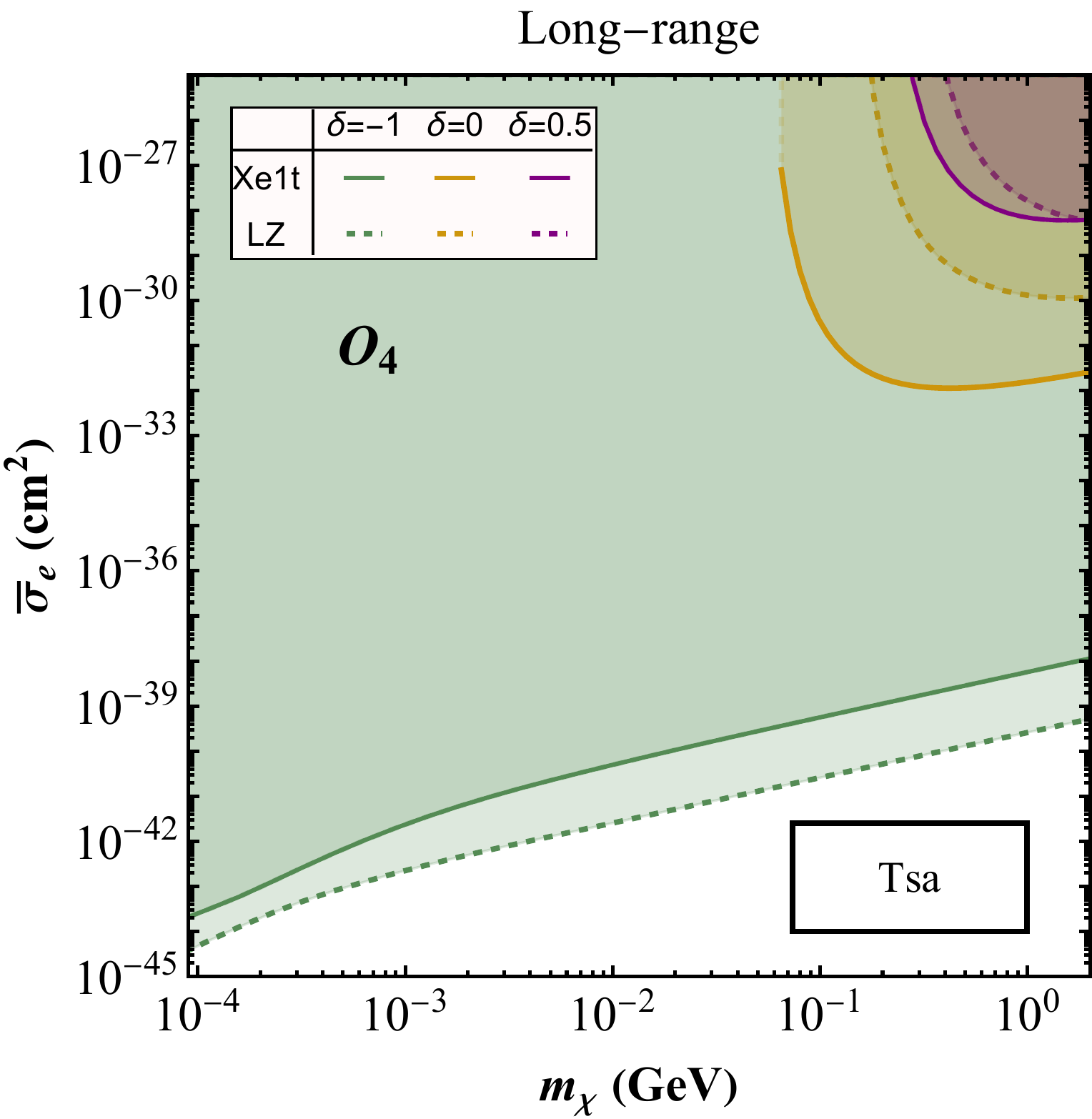}
    \includegraphics[height=6.5cm,width=5.4cm]{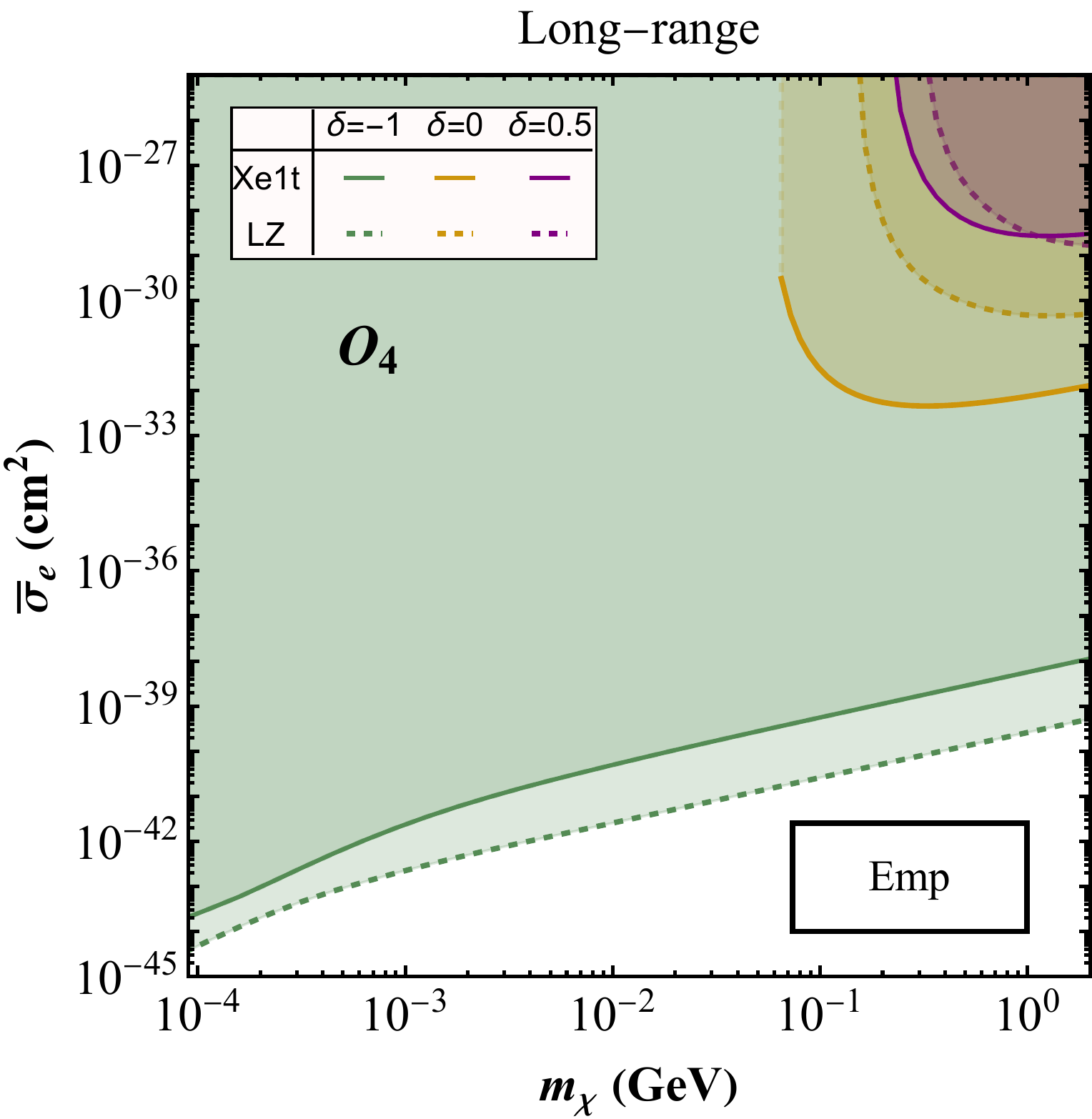}
   
    \caption{The 90\% C.L. constraints on iDM-electron cross section $\Bar{\sigma}_e$ versus DM mass $m_\chi$ from the spin-dependent contact (top) and long-range (bottom) interaction for different mass splitting $\delta$ and three velocity distribution models. We used the S2-only data from XENON1T for analysis and projected the bounds of the LZ experiment. These processes are exothermic with $\delta=-1\keV$ (green), elastic with $\delta=0\keV$ (orange), endothermic with $\delta=0.5\keV$ (purple) scattering, and include Standard Halo Model (left panel), Tsallis Mode (middle panel), Empirical Model (right panel).}
    \label{electron limited}
\end{figure}
Finally, similar previous analyses are used to give electron scattering cross sections that match the XENON1T S2-only data and to predict limits for future LZ experiment (90\%C.L.). The three different velocity distribution models are still taken into account, and we keep the exothermic (endothermic) scattering parameter of $\delta=-1\keV\,(0.5\keV)$ to demonstrate the inelastic effect on an individual effective spin-dependent operator $\boldsymbol{O_4}$ in the contact/long-range interaction, as shown in Fig.~\ref{electron limited}. 

The iDM-electron scattering bounds resembles Migdal's behaviour in Fig.~\ref{sigma}. Exothermic scattering retains more sensitivity to low mass dark matter, while endothermic scattering preserves the opposite property. This is the similarity between Migdal and electron scattering that we discussed previously, while the transfer momentum $\vec{q}$ is the crucial difference between them (reflected in the different regions of the ionization function). As mentioned previously, the effects of the velocity distribution remain slight, and only the Tsallis model differs from the other two models at higher velocity tails. Besides, we would like to emphasize that the contact and long-range interactions differ by a factor $(\frac{\alpha m_e}{q})^4$, which makes the difference between the two results quite significant. For exothermic scattering in long-range interaction, heavier DM masses ($m_\chi\gtrsim0.5\MeV$) and larger $|\delta|$ lead to a larger transfer momentum $q$, resulting in a significant relative suppression. Conversely, for $m_\chi\lesssim 0.5$ MeV, this effect is less pronounced.

\section{Conclusion}   
\label{sec6}
Although experimental work in direct dark matter detection has yielded fantastic results for exploring the DM parameter space, the future detection of sub-GeV dark matter remains a significant challenge. For sub-GeV dark matter, the electron spectrum induced by the Migdal effect and DM-electron scattering provides a detectable window for direct detection experiments near low thresholds. However, the features of inelastic dark matter and the velocity distribution functions from different dark matter halos can have a critical impact on this electron spectrum. In this paper, we consider inelastic dark matter characterised by mass splitting $\delta$ and importing the Tsallis, an Empirical and Standard Halo model of the velocity distribution function. We use a concise non-relativistic effective field theory to study the Migdal effect and electron scattering induced by inelastic dark matter through spin-dependent interaction. With data from XENON1T, we yield inelastic dark matter-nucleus Migdal/electron scattering cross sections. In the analysis of the Migdal effect, we have taken an oversimplified nuclear form factor, which makes the ``only-proton/neutron" cross section differ by about 1000. Based on our choice of astronomical parameters, the Tsallis model can have even an order of magnitude different limit on the cross section than the other two models.

We selected some currently proposed astrophysical parameters as benchmark values~\cite{Baxter:2021pqo} to compare the effects of DM halo models in different scattering processes and obtained conservative results. These results will become more transparent with the inflow of data from ongoing or upcoming DM direct detection experiments. At that time, one can compare our results with the new data to constrain the astrophysical parameters of DM particles and uncover potential DM halo models.

Finally, our work considers a single spin operator $\boldsymbol{O_4}$; more complete interaction models should be discussed. Moreover, these models may induce spin operators with velocity dependence. 
In future work, after considering the changes brought by iDM to the velocity operator $\vec{v}^{\perp}_{\mathrm{inel}}$, such as $\boldsymbol{O_7}$, $\boldsymbol{O_{12}}$ and $\boldsymbol{O_{14}}$~\cite{Barello:2014uda,Catena:2019gfa} with velocity dependence, more constraints may be imposed on the parameter space of sub-GeV DM.

\section{Acknowledgements}

We appreciate Jayden L. Newstead and James Blackman Dent for the code and helpful discussions on inelastic dark matter scattering in the EFT context. This work is supported by the National Natural Science Foundation of China (NNSFC) under grants No. 12275134, by Natural Science Foundation of Shandong Province under the grants ZR2018QA007.
\bibliography{refs}

\end{document}